\documentclass[twocolumn]{aastex6}

\begin{document}


\title{Demonstration of magnetic field tomography with starlight polarization towards a diffuse sightline of the ISM}



\author{Georgia V. Panopoulou\altaffilmark{1}, Konstantinos Tassis\altaffilmark{2,3}, Raphael Skalidis\altaffilmark{2,3}, Dmitriy Blinov\altaffilmark{3,4}}
\author{Ioannis Liodakis\altaffilmark{5}, Vasiliki Pavlidou\altaffilmark{2,3}, Stephen B. Potter\altaffilmark{6}, Anamparambu N. Ramaprakash\altaffilmark{7}}
\author{Anthony C. S. Readhead\altaffilmark{1}}
\author{Ingunn K. Wehus\altaffilmark{8}
}
\affil{1. California Institute of Technology, 1200 East California Boulevard, Pasadena, CA, 91125, MC 249-17, USA\\
2. Department of Physics and Institute for Theoretical and Computational Physics University of Crete, 70013, Heraklion, Greece\\
3. Foundation for Research and Technology – Hellas, IESL \& Institute of Astrophysics, Voutes, 70013, Heraklion, Greece\\
4. Astronomical Institute, St. Petersburg State University, Universitetsky pr. 28, Petrodvoretz, 198504 St. Petersburg, Russia\\
5. KIPAC, Stanford University, 452 Lomita Mall, Stanford, CA, 94305, USA\\
6. South African Astronomical Observatory, PO BOX 9, Observatory, 7935, Cape Town, South Africa\\
7. Inter-University Centre for Astronomy and Astrophysics, Post bag 4, Ganeshkhind, Pune, 411007, India\\
8. Institute of Theoretical Astrophysics, University of Oslo, P.O. Box 1029 Blindern, NO-0315 Oslo, Norway} 

\begin{abstract}
The availability of large datasets with stellar distance and polarization information will enable a tomographic reconstruction of the (plane-of-the-sky-projected) interstellar magnetic field in the near future. We demonstrate the feasibility of such a decomposition within a small region of the diffuse ISM. We combine measurements of starlight (R-band) linear polarization obtained using the RoboPol polarimeter with stellar distances from the second \textit{Gaia} data release. The stellar sample is brighter than 17 mag in the R band and reaches out to several kpc from the Sun. HI emission spectra reveal the existence of two distinct clouds along the line of sight. We decompose the line-of-sight-integrated stellar polarizations to obtain the mean polarization properties of the two clouds. The two clouds exhibit significant differences in terms of column density and polarization properties. Their mean plane-of-the-sky magnetic field orientation differs by 60$^\circ$. We show how our tomographic decomposition can be used to constrain our estimates of the polarizing efficiency of the clouds as well as the frequency dependence of the polarization angle of polarized dust emission. We also demonstrate a new method to constrain cloud distances based on this decomposition. Our results represent a preview of the wealth of information that can be obtained from a tomographic map of the ISM magnetic field.

\end{abstract}




\section{Introduction} \label{sec:intro}

Starlight polarization contains information on the properties of the interstellar magnetic field that lies between the star and the observer. Elements of the three-dimensional geometry of the field are encoded in the angle of the linear polarization (or polarization angle, $\theta$), and the fractional linear polarization ($p$, expressed in percentage of the total light intensity). The first observable, $\theta$, depends on the plane-of-the-sky orientation of the magnetic field and the grain alignment efficiency, and their variation along the line-of-sight. The second, $p$, depends additionally on the inclination of the field along the line-of-sight \citep{LeeDraine}. While deducing these three-dimensional properties of the field from a single stellar polarization measurement is impossible (without ample supplementary knowledge), the problem is simplified by considering ensembles of stars in conjunction with distance information. 

This potential of starlight polarization was exploited early on in the history of optical polarimetry to reconstruct the orientation of the large-scale galactic magnetic field as a function of distance from the Sun. Using polarimetry and distances for thousands of stars, \cite{lloyd1973} and \cite{fowler1974} produced maps of polarization in increments of 200 pc along the line of sight and $10^\circ-20^\circ$ on the plane of the sky. Their maps reached out to $\sim$2 kpc in the galactic plane and out to 600 pc at $|b| > 20^\circ$ and showed correlations between the polarization orientations and the local spiral arm. Their work was extended by \cite{ellisaxon1978} to include 5000 stars (within $|b| < 15^\circ$), resulting in a better statistical description of the magnetic field on scales of hundreds of parsecs. 

Subsequent studies in the optical have focused on reconstructing the properties of the magnetic field within smaller regions of space.
\cite{anderssonpotter} isolated the effect of the Southern Coalsack dark cloud on the polarization of starlight from that of foreground material, leading to a better estimation of the magnetic field strength within the cloud.
\cite{Li2006} used stellar polarizations and distances within volumes of 400 pc surrounding the Giant Molecular Cloud NGC 6334,
in order to remove the contribution of foreground/background material and isolate the magnetic field orientation local to the cloud. 
A number of works have deduced the polarizing properties of discrete clouds along sightlines toward young open clusters \citep[e.g.][]{breger1986,breger1987,vergne,eswaraiah2012}, where distances are well-known. The idea of tomographic decomposition was extended to the NIR by \citet{pavel}, who, in the absence of stellar distances, used red clump stars as standard candles.

Lack of distance information has been a major obstacle in mapping the (plane-of-the-sky) ISM magnetic field orientation in three dimensions by use of this technique. However, the advent of the \textit{Gaia} astrometric mission \citep{gaiamission}, offers an avenue to revolutionize such an exploration. With precise parallaxes for billions of stars, it will be possible to produce such a three-dimensional map of unprecedented accuracy. 

The second major obstacle has been the sparsity of existing stellar polarization measurements. Previous works provide either a high density of polarization measurements within small regions \citep[usually targeting molecular clouds, e.g. ][]{pereyra, franco} or a wider sky fraction in more diffuse regions but with highly non-uniform and sparse coverage \citep[e.g.][]{heiles,santos2011,berdyugin,cotton}.

At low galactic latitude, the Galactic Plane Infrared Polarization Survey \citep[GPIPS][]{clemens} has significantly improved coverage and density of measurements in the NIR. At high galactic latitude, future large-scale optical polarimetry surveys promise to fill the remaining gap \citep{southpol, pasiphae}. 
With survey depths two to three magnitudes fainter than the current state of the art \citep{heiles, berdyugin}, these surveys will increase the number of stellar polarizations per unit area by orders of magnitude compared to existing datasets. This will enable mapping of the plane-of-the-sky magnetic field orientation in the diffuse ISM down to the spatial scales of individual clouds, matching existing datasets in dense molecular clouds \citep[e.g.][]{marchwinski}. By combining such measurements with stellar distances, these surveys will open up new paths to explore the properties of the plane-of-the-sky component of the magnetic field along the third (line-of-sight) dimension. In conjunction with complementary measurements of the line-of-sight component of the magnetic field \citep[obtained through Faraday tomography, e.g.][]{heald,ferriere}, as well as line-of-sight-integrated measures \citep[e.g. thermal dust emission and synchrotron polarization,][]{planckmodel} such information can aid in ongoing efforts to model the three-dimensional Galactic magnetic field \citep[e.g.][]{imagine}.

In this work we wish to demonstrate some of the techniques needed for such a tomographic survey by observing a small region of the intermediate Galactic latitude sky.
This work is a path-finder for the Polar Areas Stellar Imaging in Polarization High Accuracy Experiment \citep[PASIPHAE,][]{pasiphae} as it reaches comparable depth and polarimetric accuracy as expected for PASIPHAE for stars in the direction of a carefully chosen diffuse region. 

We begin by exploring the properties of the ISM in the selected region with the help of HI spectral information (Section \ref{sec:HIdust}). We then describe our observations of optical polarization (Section \ref{sec:obs}), followed by the data reduction and calibration (Section \ref{sec:reduction}). We analyse the properties of the measured polarizations as projected on the plane of the sky and as a function of distance, by making use of the recent \textit{Gaia} second data release (DR2) in section \ref{sec:results}. This information allows us to decompose the plane-of-the-sky magnetic field orientation along the line of sight. We discuss our findings in Section \ref{sec:discussion} and conclude in Section \ref{sec:conclusions}.

\begin{figure*}
\centering
\includegraphics[scale=1]{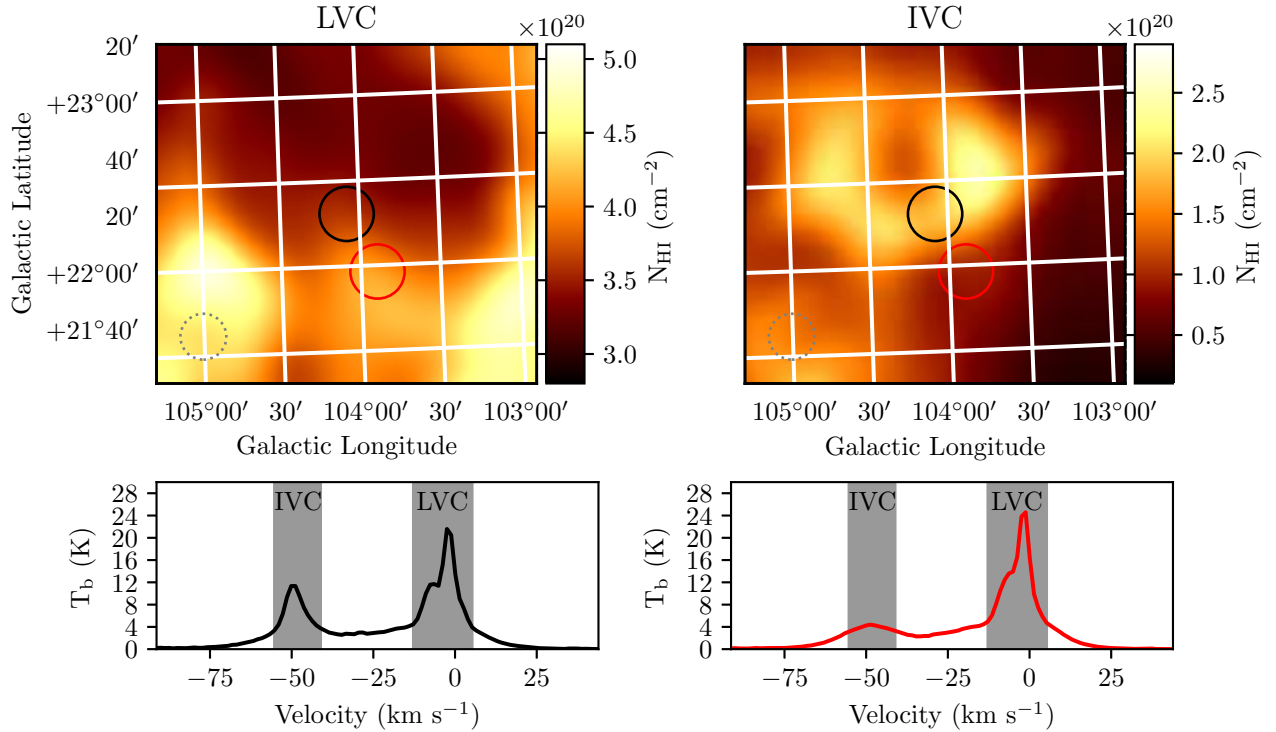}
\caption{HI emission in the surveyed region from the HI4PI dataset. Top panels: 2-Cloud (black circle) and 1-Cloud (red circle) regions are shown on the plane of the sky. The background image shows the HI column density within the range of velocities of the LVC (left) and that of the IVC (right). The gray circle in the bottom left corner shows the beam (FWHM) of the HI map. Bottom panels: Average spectrum in the 2-Cloud region (left panel) and the 1-Cloud region (right panel). The HI spectrum shows two very distinct components around -50 km $\rm{s^{-1}}$ (IVC) and -2.5 km $\rm{s^{-1}}$ (LVC). The range of velocities of each component is marked with a gray band. Velocities are with respect to the Local Standard of Rest (LSR).}
\label{fig:HIspec}
\end{figure*}

\section{The distribution of gas along the line of sight}
\label{sec:HIdust}

Since stellar polarization is imparted through dichroic extinction of light by dust grains (which are aligned with the magnetic field), 
this observable preferentially traces the magnetic field in the neutral atomic and molecular phase of the ISM, which dominate the dust
column. 

Dust and HI are tightly correlated in the diffuse ISM \citep[e.g.][]{Bohlin}. We use the kinematic information from HI line emission spectra to infer properties of the distribution of atomic gas (and consequently, dust) along the line-of-sight. To this end, we employ the publicly available spectral cube from the HI4PI survey \citep{HI4PI} which contains the selected region. The region is defined as a circle of radius 0.16$^\circ$ centered on (l,b) = (104.08$^\circ$, 22.31$^\circ$). Figure \ref{fig:HIspec} (bottom left) shows the HI spectrum averaged within this area. The spectrum reveals the existence of two kinematically distinct components of HI emission. One is located around a velocity of -2.5 km $\rm{s^{-1}}$ and has a peak brightness temperature ($\rm T_b$) of 22 K. The other has a much lower peak $\rm T_b $ of 12 K and is located at -50 km $\rm{s^{-1}}$. This double-peaked spectrum, with components that are well separated in velocity, implies that the neutral ISM mass is distributed in at least two spatially distinct components along the line of sight. 

The very small velocity (compared to the local standard of rest) of the HI component that peaks at -2.5 km $\rm{s^{-1}}$ suggests that the emission originates nearby. We shall refer to this component as the Low Velocity Cloud (LVC). The second component is at velocities consistent with the class of Intermediate Velocity Clouds \citep{wakker} and we shall refer to it as the IVC. 

The contribution of the two components to the total atomic gas content of the target region is uneven, with the LVC clearly dominating the emission. We calculate the HI column density of each component, $\rm N_{HI}$, using: ${\rm{N_{HI}}} = \int_{v=v_{min}}^{v_{max}} 1.823 \times 10^{18} {\rm{T_b}}(v)  dv \,\,\,\, \rm cm^{-2}/(K \, km \,s^{-1})$, where $\rm{T_b}$ is the brightness temperature of the HI emission in K, $dv$ is the spectral resolution of the HI4PI data (1.288 km $\rm{s^{-1}}$) and the summation takes place over the range of velocities [$v_{min},v_{max}$] within which each component dominates. This follows from the equations of radiative transfer for the HI line under the assumption of optically thin emission \citep[e.g.][]{kulkarni}. We define a threshold of $\rm{T}_b$ at 4 K, which separates the spectrum into the two components, and integrate the emission within the velocities where $\rm{T}_b > 4$ K: -55 to -41 km $\rm{s^{-1}}$ for the IVC and -12 to 5 km $\rm{s^{-1}}$ for the LVC (these ranges are shown with shaded gray regions in the spectrum of Figure \ref{fig:HIspec}). We find that the HI column density of the LVC is a factor of $\sim$2 higher than that of the IVC ($\rm N^{LVC}_{HI} = 3.5\times 10^{20} \, cm^{-2}$ and $\rm N^{IVC}_{HI}=1.8\times 10^{20} \, cm^{-2}$).

The two clouds are not only different in terms of their total (atomic) gas content, but they also show distinct morphologies on the plane of the sky. The top panels of Figure \ref{fig:HIspec} show maps of $\rm N_{HI}$ inferred from integrating the emission within $\sim$1$^\circ$ from our target region over the velocity range where the IVC dominates (-55 to -41 km $\rm{s^{-1}}$, left panel) and where the LVC dominates (-12 to 5 km $\rm{s^{-1}}$, right panel). The IVC has a bubble-like shape with a well-defined boundary towards the south-east. In contrast, the LVC is much more spread out and exhibits less abrupt spatial variations. 

These characteristics of the two clouds allow us to define a `control' region for our experiment, marked with a red circle in the top panels of Figure \ref{fig:HIspec}. This region is identical in size to the target region but is centered on a neighbouring position where the IVC emission is suppressed: (l,b) = (103.90$^\circ$, 21.97$^\circ$). This can be seen by inspecting the spectrum within this control region (bottom right panel of Figure \ref{fig:HIspec}). The $\rm N_{HI}$ of the IVC here is a factor of two lower compared to that in the target region. We therefore expect that in the target region, both clouds will contribute to the stellar polarizations, while in the control region, the effect of the IVC on starlight polarization will be minimal. Measurements in the control region can thus be used to isolate the effect of the LVC.

In the following we will refer to the region with significant contribution from the IVC and LVC as the 2-Cloud region (black circle in Fig. 1 top panels) and to that with mainly LVC emission as the 1-Cloud region (red circle in Fig. 1 top panels).

\section{Polarimetric Observations} \label{sec:obs}

\begin{figure}
\centering
\includegraphics[scale=1]{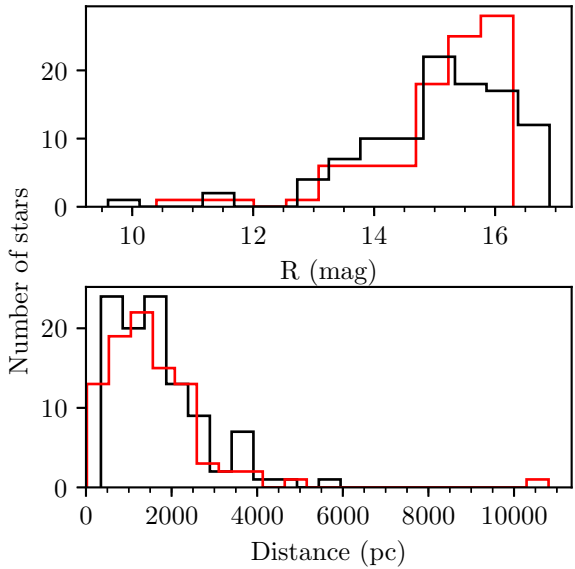}
\caption{Properties of the observed stellar samples. Top: Distributions of R magnitudes from the USNO-B1 catalogue. Bottom: Distributions of the maximum-likelihood distances from the catalogue of \citet{bailer-jones}. In both panels, the red line corresponds to stars in the 1-Cloud sample, and the black line to stars in the 2-Cloud sample.}
\label{fig:rmags}
\end{figure}

We selected a sample of stars in each region, with R $<$ 17 mag, from the USNO-B1 catalogue \citep{usnob}. Our sample consists of 196 stars in total (103 stars in the 2-Cloud region and 93 stars in the 1-Cloud region), with 9.6 mag $<$ R $<$ 16.9 mag. The distributions of R magnitudes (from USNO-B1) for the 1-Cloud and 2-Cloud samples are shown in the top panel of Figure \ref{fig:rmags}. The samples are not photometrically complete in these regions due to time constraints. 

We obtain stellar distances by cross-matching our targets with the catalogue of \citet{bailer-jones}, who provide a probabilistic estimate of the distance to stars in \textit{Gaia} DR2 \citep{gaiadr2}. They infer the posterior probability density function (PDF) of the distance, given the measured parallax, using an exponentially decreasing space density prior. The catalogue presents the mode of the posterior PDF for the distance and we will refer to this value as the distance to the star. Uncertainties are provided as the (asymmetric) bounds of the highest density interval (equivalent to $\pm$1$\sigma$ for a Gaussian distribution). 

The bottom panel of Figure \ref{fig:rmags} shows the distributions of stellar distances in our samples from this catalogue. Three of our sources (one with R = 13.4 mag and two with R = 15.1 mag) have undefined distances in the \citet{bailer-jones} catalogue and are not included in the distribution. We find that the photometric depth of our survey is sufficient to cover a wide range of distances and that there is no significant difference between the distribution of stellar distances in the two regions (a two-sample, two-sided, K-S test reports a p-value of 0.8).

We performed polarimetric observations of our sample during 2016, 2017, and 2018 with the RoboPol polarimeter \citep{ramaprakash}, which is mounted on the 1.3 m Ritchey-Chr\'etien telescope at the Skinakas Observatory in Crete, Greece. The instrument is an imaging polarimeter, which uses two half-wave plates and two Wollaston prisms to simultaneously measure the relative Stokes parameters $q = Q/I$ and $u = U/I$ (I is the total intensity and Q, U are the absolute Stokes parameters). Observations were conducted during 13 nights from May to July 2016, during 5 nights in July 2017, and during 6 nights in August 2018. Observing time was shared with other projects. The observing time for science targets was about 66 hours in total. 

Our strategy was to place each star in the central region of the instrument. In this region, a mask reduces the sky background compared to the rest of the field of view \citep[see Fig. 4 of][]{King2014}. The instrumental systematic uncertainty is below 0.1\% in $q$ and $u$ within this area \citep{Skalidis2018}, while in the entire field of view ($13.6\arcmin \times 13.6 \arcmin$) this increases by a factor of three \citep{Panopoulou2015}. The exposure time for each target was set with the aim of obtaining significant measurements of stellar polarization. The median exposure time per source was 14 minutes, while only 5 sources required more than 50-minute exposures each. We use a single-epoch observation of each source for our analysis (measurements from a single consecutive series of exposures in the mask, taken on the same night).

\section{Polarization Data reduction \& calibration} \label{sec:reduction}

The data are reduced using the RoboPol pipeline \citep{King2014}. The pipeline measures the relative Stokes parameters $q$ and $u$ for each target through differential aperture photometry. We use the version of the code described in \citet{Panopoulou2015}, which optimizes the aperture size for each source. By default, the pipeline corrects the Stokes parameters according to a model of the instrumental polarization  \citep[described in][]{King2014}. We turn this option off when processing the data in order to avoid unknown uncertainties that may arise from the modelling. Instead, we correct for instrumental polarization directly using measurements of polarization standard stars placed in the mask (where our target stars were also placed). 

\begin{figure}
\centering
\includegraphics[scale=1]{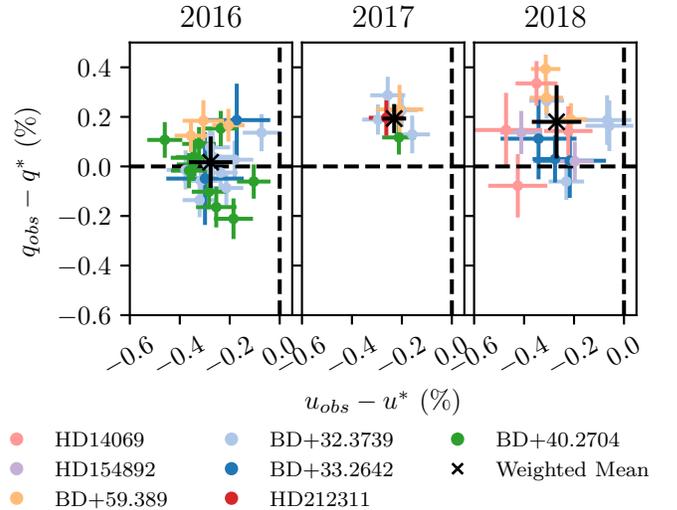}
\caption{Residuals of the observed relative Stokes parameters (in the R-band) of polarization calibrator stars from their literature values for the three observing runs (2016, 2017, and 2018 shown in the left, middle, and right panels, respectively). A different color is used to mark each calibrator star. Measurement uncertainties are purely statistical (from photon noise error propagation). A black cross marks the weighted mean of the measurements (mean instrumental polarization) for each run, with corresponding error bars marking the standard deviation (systematic uncertainty). }
\label{fig:stands}
\end{figure}

We find the differences of the observed relative Stokes parameters ($q_{obs},u_{obs}$) of our calibrators from their (true) literature values ($q^*,u^*$). 
These differences (residuals) are shown in Figure \ref{fig:stands}. We only use measurements of standards observed on the same nights as the project targets were observed. Because the 2016 observing run was longer, there are significantly more measurements that can be used for calibration for this run (28 - left panel) compared to the 2017 (7 - middle panel) and 2018 (16 - right panel) runs. The literature values of the standard stars are shown in Table \ref{tab:stands}. 

\begin{table}
    \centering          
    \caption{Literature polarization of standard stars used for the instrument calibration.}                  
    \begin{tabular}{c c c c l l l }   
      \hline\hline                     
       Name& $p$(\%) & $\theta$ & Band & Ref. \\ 
       \hline  
       BD +32 3739 	& 0.025$\pm$0.017	& 35.79$^{\circ}$		& V &	1	\\
       BD +33 2642   & 0.20$\pm$0.15  & 78$^{\circ}\pm$ 20$^{\circ}$  &R & 2\\       
       BD +40 2704   & $0.07\pm0.02$& 57$^{\circ}\pm 9^{\circ}$& ?& 3\\
       BD +59 389 	& 6.430$\pm$0.022	& 98.14$^{\circ}$ $\pm$ 0.10$^{\circ}$& R &	1	\\ 
       HD 14069	    & 0.022 $\pm$ 0.019  & 156.57$^{\circ}$   		&  V & 1\\
       HD 154892    & 0.05 $\pm$ 0.03    & --                & B  & 4\\
       HD 212311 	& 0.034 $\pm$ 0.021 	& 50.99$^{\circ}$   	&  V & 1\\
       \hline          
    \end{tabular} 
    References. (1)\citet{schmidt}; (2) \citet{Skalidis2018}; (3) \citet{Berdyugin2002}; (4) \citet{Turnshek}
    \label{tab:stands}
  \end{table} 

We find the weighted mean $q$ and $u$ for each run, which is our best estimate for the level of instrumental polarization ($q_{inst}$, $u_{inst}$). We assign the standard deviation of the measurements to be the uncertainty on this value (systematic uncertainty, $\sigma_{q,inst}$, $\sigma_{u,inst}$). The standard deviation most likely overestimates the systematic uncertainty of the instrument, but it is a conservative estimator compared to the more commonly used standard error of the mean. The instrumental polarization varies slightly between the three observing seasons. The values for the instrumental polarization for 2016 are: $q_{inst} = -0.01 \pm 0.13$\%, $u_{inst} = -0.28 \pm 0.08$\%, for 2017: $q_{inst} = 0.19 \pm 0.06$\%, $u_{inst} = -0.23 \pm 0.05$\%, and for 2018: $q_{inst} = 0.18 \pm 0.15$\%, $u_{inst} = -0.27 \pm 0.10$\%. The variations in instrumental polarization between different years are due to the routine removal of the instrument from the telescope mount at the end of an observing season (November) and its reinstallation at the beginning of the next season (April).

Measurements of our target stars are corrected for the instrumental polarization by subtracting the weighted mean $q_{inst}$, $u_{inst}$ (determined for its corresponding observing run) from the observed value of $q$ and $u$ and propagating the systematic uncertainty to the final result. Our measurements have not been corrected for the rotation of the instrument frame with respect to the celestial reference frame. This rotation has been measured using polarized standards in all observing seasons and was found to be $ < 1^\circ$, which is less than the typical $1\sigma$ uncertainty of our measurements ($5^\circ$). 

The fractional linear polarization, $p$ is calculated from the Stokes parameters through:
  \begin{equation}
    p = \sqrt{q^{2} + u^{2}},\hspace{2mm} \sigma_{p} = \sqrt{\frac{q^{2}
	\sigma^{2}_{q}+u^{2}\sigma^{2}_{u}}{q^{2}+u^{2}}} 
    \label{eq:polarization_stokes}
  \end{equation}    
where the uncertainties on the Stokes parameters $\sigma_q$ and $\sigma_u$ include both statistical and systematic uncertainties.

As $p$ is a biased estimator of the true fractional linear polarization, $p_0$, we correct for this bias using the estimator proposed by \citet{plaszczynski}:
\begin{equation}
p_d = p - \sigma_p^2 \frac{1-e^{-p^2/\sigma_p^2}}{2p}
\label{eq:pd}
\end{equation}
and calculate the 68\%, 95\% and 99\% confidence intervals on $p_0/\sigma_p$ through the provided analytical expressions (equations 26 in their paper). This estimator is superior in correcting for the bias in the low signal-to-noise ratio in $p$ regime \citep{plaszczynski,montier} compared to the most commonly used estimator discussed in \citet{vaillancourt}.

For measurements with $\sigma_q \approx \sigma_u$ (as is the case in our work), the polarization angle found through\footnote{The polarization angle $\theta$ is calculated using the two-argument arctangent to lift the $\pi$ ambiguity. It is measured with respect to the International Celestial Reference Frame (ICRS) according to the IAU convention.}:
\begin{equation}
    \theta=\frac{1}{2}\arctan\left(\frac{u}{q}\right)
    \label{eq:angles_stokes}
   \end{equation}
is an unbiased estimator of the true $\theta_0$ \citep{montiera}. We determine the uncertainty in $\theta$, $\sigma_\theta$, following \citet{Naghizadeh1993}. We solve the integral:
\begin{equation}
 \int_{-1\sigma_\theta}^{1\sigma_\theta} G(\theta;P_0) d\theta = 68.27\%,
 \label{eqn:sigmatheta}
\end{equation}
where $P_0 = p_0/\sigma_p$ and $G$ is the probability density function defined as:
\begin{equation}
 G(\theta;\theta_0;P_0) = \frac{1}{\sqrt{\pi}} \left\{ \frac{1}{\sqrt{\pi}} + \eta_0 e^{\eta_0^2}
 [ 1 + erf(\eta_0) ] \right\} e^{-\frac{P_0^2}{2}},
\end{equation}
where $\eta_0 = P_0/\sqrt{2}\cos{2(\theta - \theta_0)}$ and $erf$ is the 
Gaussian error function.

\begin{figure*}
\centering
\includegraphics[scale=1]{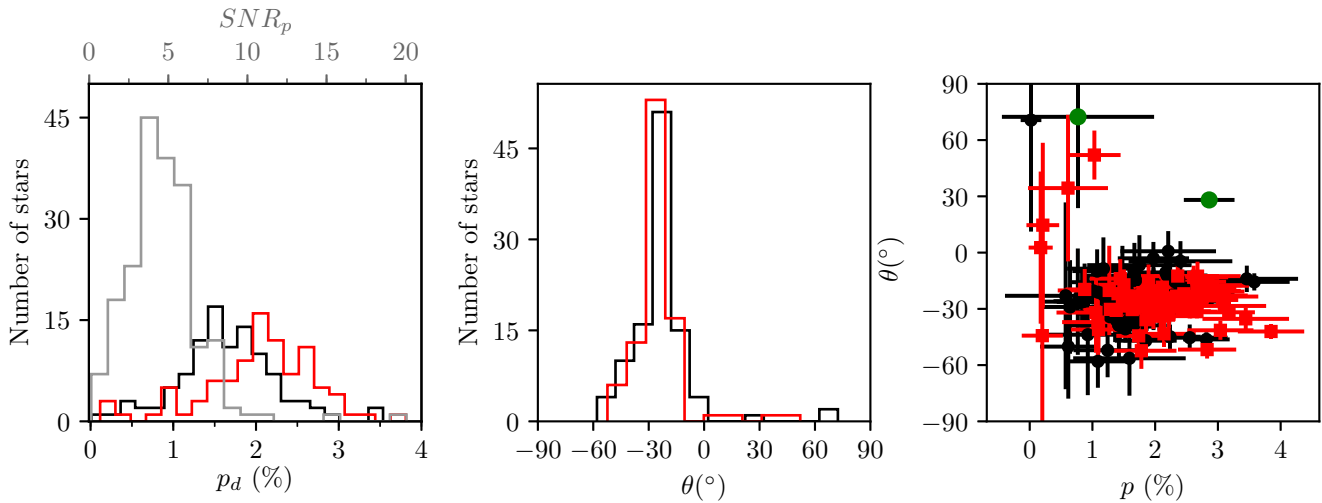}
\caption{Measured stellar polarizations. Left: Distribution of debiased fractional linear polarizations ($p_d$) (bottom axis) for the stars in the 2-Cloud region (black line) and 1-Cloud region (red line), as well as $p/\sigma_p$ for all stars (gray line, top axis). The errors $\sigma_p$ contain both statistical and systematic uncertainties. Middle: Distribution of polarization angles, $\theta$, in the 2-Cloud region (black) and 1-Cloud region (red). Right: $\theta$ versus $p$ for sources that lie within the 2-Cloud region (black circles) and for those in the 1-Cloud region (red squares). The green points mark the outliers defined in section \ref{subsec:stellarp}.}
\label{fig:ITqu}
\end{figure*}

\section{Results} 
\label{sec:results}

\begin{table*}
\centering
\caption{Catalogue of stellar polarization measurements (full table online). Columns contain: star identification number in the Gaia catalogue, star identification number in the USNO-B1 catalogue, R.A.(J2000), Dec (J2000) (from USNO-B1 catalogue), Stokes $q$ and 1$\sigma$ uncertainty, Stokes $u$ and 1$\sigma$ uncertainty, fractional linear polarization $p$ and 1$\sigma$ uncertainty, debiased fractional linear polarization $p_d$, polarization angle $\theta$ and 1$\sigma$ uncertainty, stellar distance from \citet{bailer-jones} $d$, lower and upper limits on distance ($d_{low}, d_{high}$), and flag specifying the region in which the star lies (1 for 1-Cloud region, 2 for 2-Cloud region). Intrinsically polarized candidates are flagged 0.}
\begin{tabular}{|c|c|c|c|c|c|c|c|c|c}
\hline
Gaia ID & USNO-B1 ID & R.A. ($^\circ$) & Dec ($^\circ$) & sign($q$) & $q$ & $\sigma_q$ & sign($u$)& $u$ & $\sigma_u$  \\ 
\hline
\hline
2263930248734795264 & 1620-0140825 & 294.78657 & 72.08941 & + & 0.00397 & 0.00820 & -& 0.00413 & 0.01086   \\
2263906231277703424 & 1618-0137841 & 295.37764 & 71.84215 & + & 0.00975 & 0.00421 & -& 0.01561 & 0.00355    \\ 
\hline
\end{tabular}
\begin{tabular}{c|c|c|c|c|c|c|c|c|c|}
\hline
$p$ & $\sigma_p$ & $p_d$ & sign($\theta$) & $\theta (^\circ)$ & $\sigma_\theta (^\circ)$ & $d$ (pc) & $d_{low}$ & $d_{high}$ & flag\\ 
\hline
0.00573 & 0.00967 & 0.00331 & - & 23.0 & 49.8 & 2049 & 1986 & 2116 & 2 \\ 
0.01841 & 0.00375 & 0.01803 & - & 29.0 & 06.0 & 1175 & 1151 & 1200 & 1 \\
\hline
\end{tabular}
\label{tab:data}
\tablecomments{ The table lists only the first 2 sources. It is published in its entirety in machine-readable format. A portion is shown here for guidance regarding its form and content.}
\end{table*}

\begin{figure}
\centering
\includegraphics[scale=0.7]{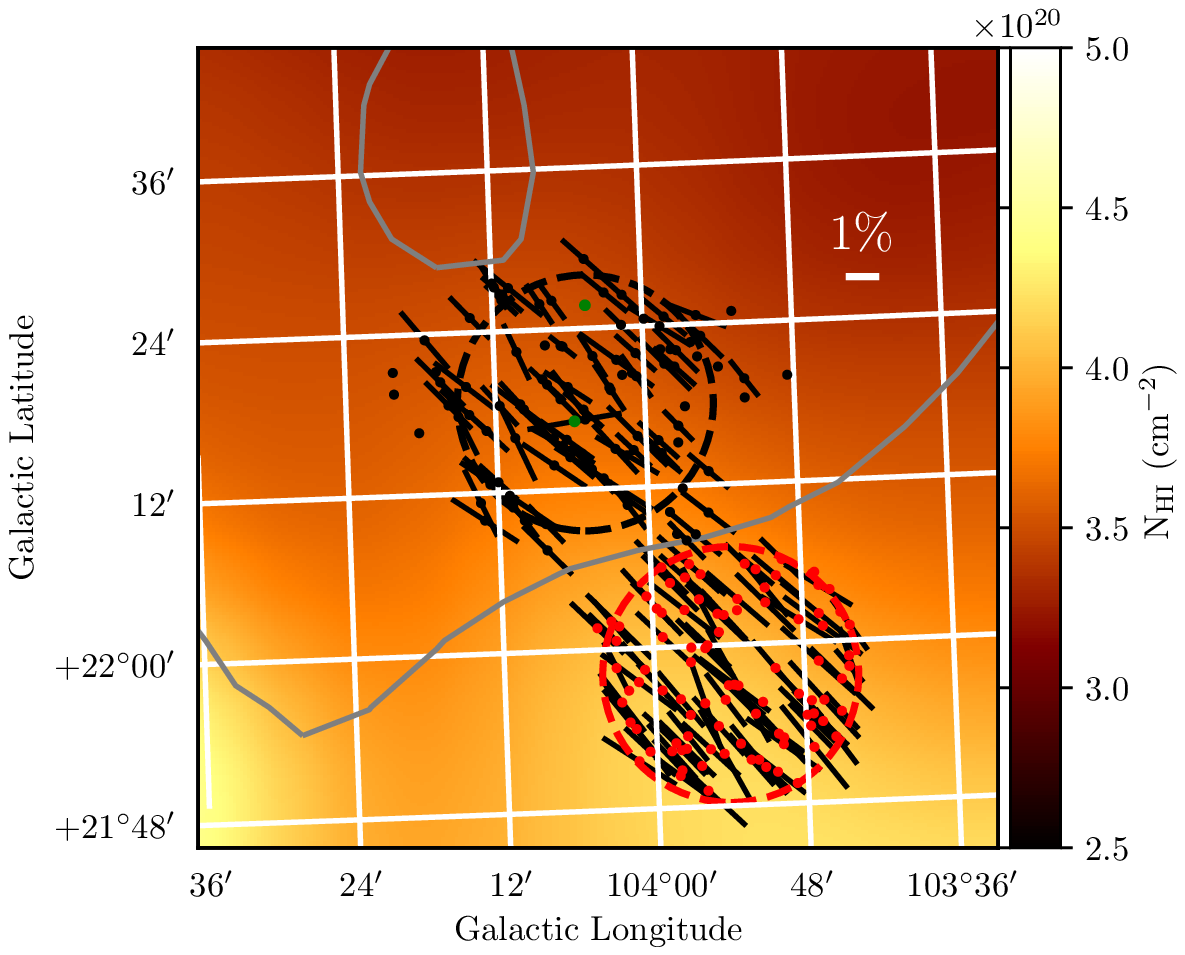}
\caption{Stellar polarization measurements overplotted on the $\rm N_{HI}$ map of the LVC. The position of each star is marked with a dot (black: 2-Cloud region, red: 1-Cloud region). For each star, a line segment is drawn that forms an angle $\theta_{gal}$ with respect to the Galactic North (increasing towards the left). We do not show segments for stars with $p/\sigma_p < 3$, which have large uncertainties in $\theta_{gal}$. The length of the segments is proportional to $p$. A line of $p=1\%$ is shown on the top right corner for scale (white segment). The gray contour outlines the emission of the IVC, at a level of $\rm N_{HI} = 1.35 \times 10^{20} \, cm^{-2}$. The large dashed open circles mark the same regions as the circles in Figure \ref{fig:HIspec}. The green points mark the outliers defined in section \ref{subsec:stellarp}}
\label{fig:segs}
\end{figure}

\subsection{Stellar polarizations}
\label{subsec:stellarp}
All polarization measurements are available in the online table accompanying the paper. We present the first two rows in Table \ref{tab:data}. We investigate the statistical properties of the measurements in Figure \ref{fig:ITqu}. The distribution of $p/\sigma_p$ (left panel, gray line) shows that the majority of our measurements are significant detections, with 78\% of the values lying above a ${\rm SNR}_{\rm p}$ of 3. The $\sigma_p$ distribution (not shown) has a mean of 0.46\% and a standard deviation of 0.17\%. Hence, our measurements are photon-noise-limited, as the systematic uncertainty is at the much lower level of 0.1\% (section \ref{sec:reduction}). Only two sources have uncertainties for which the systematic uncertainty has a significant contribution (their quoted uncertainty is $\sigma_p < 0.14\%$). 

\begin{figure*}
\centering
\includegraphics[scale=1]{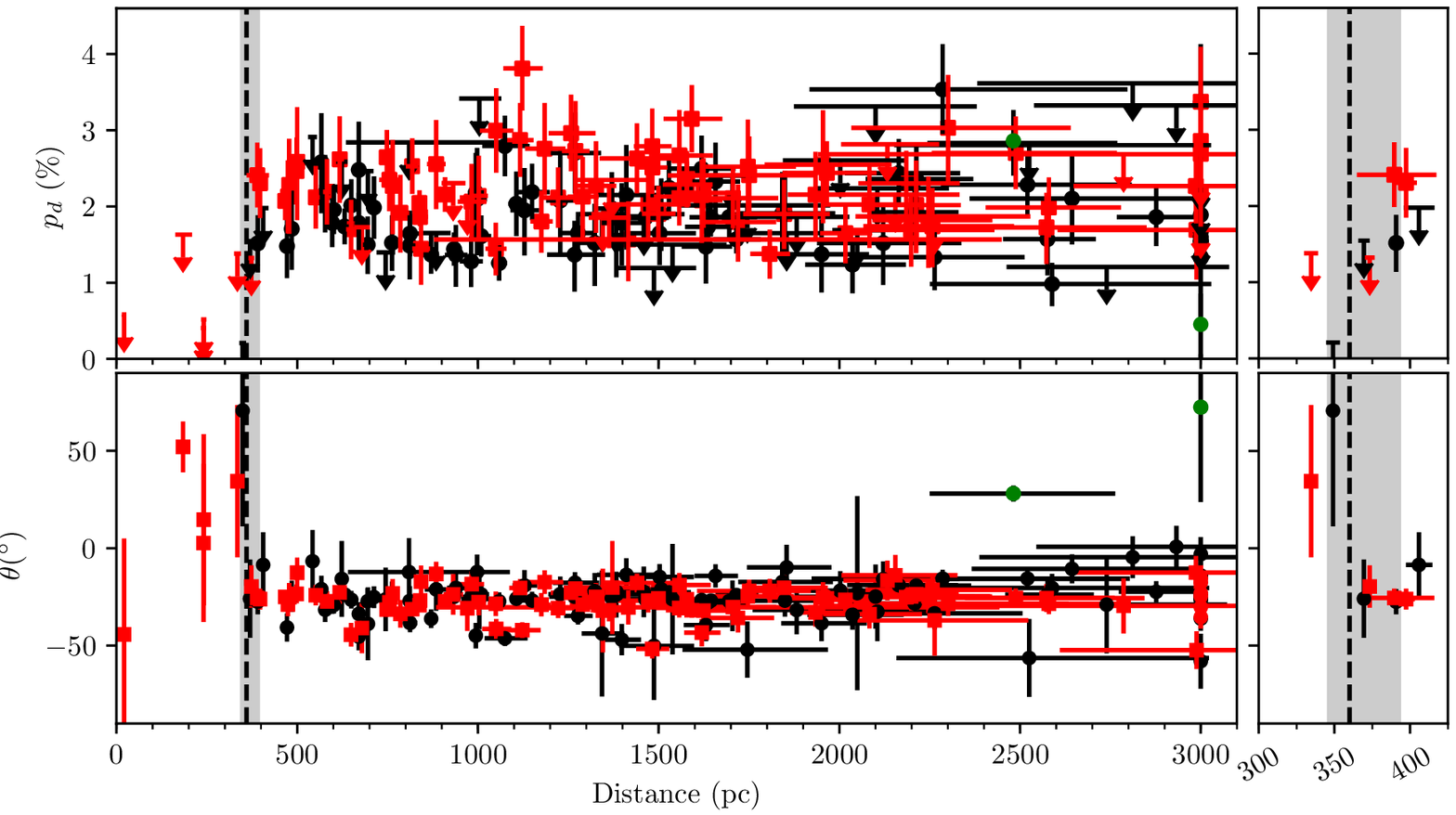}
\caption{Fractional linear polarization debiased using equation \ref{eq:pd} ($p_d$), top, and polarization angle ($\theta$), bottom, versus distance for the stars in our sample.  Symbols as in Figure \ref{fig:ITqu}. Stars with distances farther than 3 kpc have been shifted to 3 kpc for better visualization. The dashed vertical line marks our estimate for the distance of the LVC (360 pc), while the gray band marks the range of possible LVC distances. Three of the 196 sources are not shown because they have undefined distances. The green points mark the outliers defined in section \ref{subsec:stellarp}. In the top panel, 2$\sigma$ upper limits are shown for measurements with $p_d/\sigma_p < 3$. The insets on the right show a zoomed-in version of the main panels ($p_d$, top, $\theta$, bottom), within the range 300 - 425 pc.}
\label{fig:ptdist}
\end{figure*}

The distribution of $p_d$ in the 2-Cloud region (Figure \ref{fig:ITqu}, left panel, black line) has a mean value of 1.6\%, which is slightly less than that found in the 1-Cloud region (1.9\%, Figure \ref{fig:ITqu}, left panel, red line). A two-sided K-S test rejects the null hypothesis that the two distributions arise from the same parent distribution, with a p-value of $\sim10^{-7}$.
The distribution of $\theta$ (Figure \ref{fig:ITqu}, middle panel) is strongly peaked in both regions with a standard deviation of 17$^\circ$ and 14$^\circ$ in the 2- and 1- Cloud regions, respectively, and a mean of $\sim$-25$^\circ$. The mean $\theta$ differs by only 2$^\circ$ between the two regions. The two distributions of $\theta$ are not significantly different (the two-sample K-S test p-value is 0.6). There are 5 outliers which lie farther than 3 standard deviations from the mean and are easily identifiable as a tail towards large angles. Such significantly divergent measurements may arise if some subset of these sources is tracing a different fraction of the total column (e.g. they may be foreground to the clouds) and/or if there is intrinsic polarization associated with some of the sources. 

The $p$-$\theta$ plane (Figure \ref{fig:ITqu}, right panel) enables more detailed inspection of the characteristics of our measurements. Sources in the 2-Cloud and 1-Cloud regions are marked separately (black circles and red squares, respectively). The majority of measurements are clustered at high $p$ and negative $\theta$. There are seven sources which clearly deviate from the bulk of the points (all at $\theta>$  0$^\circ$). Of these sources, only one is a significant detection ($p/\sigma_p > 3$). The two sources marked with green lie at distances farther than 2 kpc, while the remaining sources are all nearby (within 360 pc) and are foreground sources (Section \ref{sec:dist}).

The deviant $\theta$ of the two distant sources (green points in Figure \ref{fig:ITqu}, right panel) may be a sign of intrinsic polarization. 
We could not find auxiliary evidence of intrinsic polarization for either source (USNO-B1 ID: 1622-0145399, R.A. = 294.55244$^\circ$, Dec: 72.23634$^\circ$,  and USNO-B1 ID: 1622-0145176, R.A. = 294.08869$^\circ$, Dec = 72.26715$^\circ$, J2000). 
We do not use these sources in the subsequent analysis.

Figure \ref{fig:segs} shows the measurements on the plane of the sky. The background image is the $\rm N_{HI}$ of the LVC and the gray contour marks the edge of the IVC, defined at a level of $\rm N_{HI} = 1.35 \times 10^{20} \, cm^{-2}$. To increase the number of measurements in the 2-Cloud sample, we observed some stars that lay slightly outside the region marked with the black circle. All stars that lie within the IVC contour are assigned to the 2-Cloud region (shown as black dots), while those that lie outside it are assigned to the 1-Cloud region (red dots). The linear segments (for all stars $p/\sigma_p > 3$) form an angle $\theta_{gal}$ compared to the Galactic reference frame\footnote{We convert polarization angle $\theta$, measured in the ICRS, to polarization angle $\theta_{gal}$, measured in the Galactic frame, following \citet{appenzeller}.}. As expected from the distributions of $\theta$ (Figure \ref{fig:ITqu}, middle) the measured polarization angles form an ordered pattern with no apparent difference between the 1-Cloud and 2-Cloud regions. This is consistent with our expectation that the LVC is dominating the signal in polarization, as is the case in HI emission (Section \ref{sec:HIdust}).

\subsection{Stellar polarization versus distance} 
\label{sec:dist}

Though stellar polarizations do not show marked statistical differences as a function of position on the plane of the sky, the situation may change by adding the information of stellar distance. Figure \ref{fig:ptdist} shows the debiased fractional linear polarization, $p_d$, (top) and polarization angle, $\theta$, (bottom) versus the maximum likelihood stellar distance from the catalogue of \citet{bailer-jones}. Stars at large distances (17 in total) are shown at a distance of 3 kpc without their distance uncertainties to facilitate visualization. The eight stars nearest to the Sun are not significantly detected in polarization ($p/\sigma_p < 3$). At farther distances we find a systematic change in both the $p$ and $\theta$ of stars. The values of $p$ are systematically higher and those of $\theta$ cluster around -24$^\circ$. This behaviour reflects the effect of the nearest cloud, the LVC. This abrupt change allows us to pinpoint the distance to the cloud with relatively high accuracy. 

The sixth nearest star, which is at a distance of 346-352 pc and is clearly unpolarized, sets a lower bound on the distance to the LVC at 346 pc. Though the seventh nearest star has $p/\sigma_p = 1.8$, its $\theta$ seems to agree with that of further away stars. We cannot be certain that it is background to the cloud. It lies at a distance of 367-372 pc. The ninth nearest star is significantly polarized ($p/\sigma_p = 6$) and lies at 366-416 pc and the tenth nearest star lies at 387-393 pc. The LVC cannot lie farther than $\sim$ 400 pc, otherwise these two stars should also be unpolarized. Therefore the distance to the LVC is determined to be within $d_{LVC} = 346 - 393$ pc. For the remainder of this work we take the coincidence of the seventh nearest star's $\theta$ with the rest of the polarization angles as evidence that it is background to the cloud. We therefore adopt a distance of 360 pc as the nominal distance to the cloud. 

At larger distances, there is no apparent shift in the properties of either $p$ or $\theta$ in the 2-Cloud region. This is consistent with our expectation that the LVC will dominate the polarization properties in the region, thus making it very difficult to discern an effect of the IVC on the line-of-sight-averaged polarization. The 1-Cloud region data also do not show any features with distance, as expected.

\subsection{The polarizing properties of each cloud}
\label{sec:tomography}

Having a precise distance to the nearby cloud (Section \ref{sec:dist}) we proceed to disentangle the effect of the two clouds on the measured polarization signal. This cannot be done on a star-by-star basis, so we consider ensembles of stars to infer the average polarization properties due to each cloud. 

The task of decomposing the polarization properties along the line-of-sight is greatly facilitated by the fact that the optical depth is small for measurements in the optical towards the diffuse ISM and hence the resulting interstellar-induced $p$ is small, typically $ << 10 \%$). In this limit of low polarization, the Stokes parameters $q$ and $u$ are additive. Suppose that two clouds\footnote{We will refer to a polarizing medium with well-defined mean magnetic field orientation and polarizing efficiency as a `cloud'.} exist along the line-of-sight, cloud A lies further from the observer than cloud B. Clouds A and B induce polarization (on unpolarized light passing through a specific position of the cloud) described by Stokes parameters $q_A,u_A$ and $q_B, u_B$, respectively. Then, a light beam that is transmitted through cloud A and subsequently through cloud B acquires a final polarization described by $q_A+q_B, u_A+u_B$ \citep[for a more detailed analysis of the equations leading to this conclusion, see e.g. the appendix of][]{Patat2010}. 

\begin{figure*}
\centering
\includegraphics[scale=1]{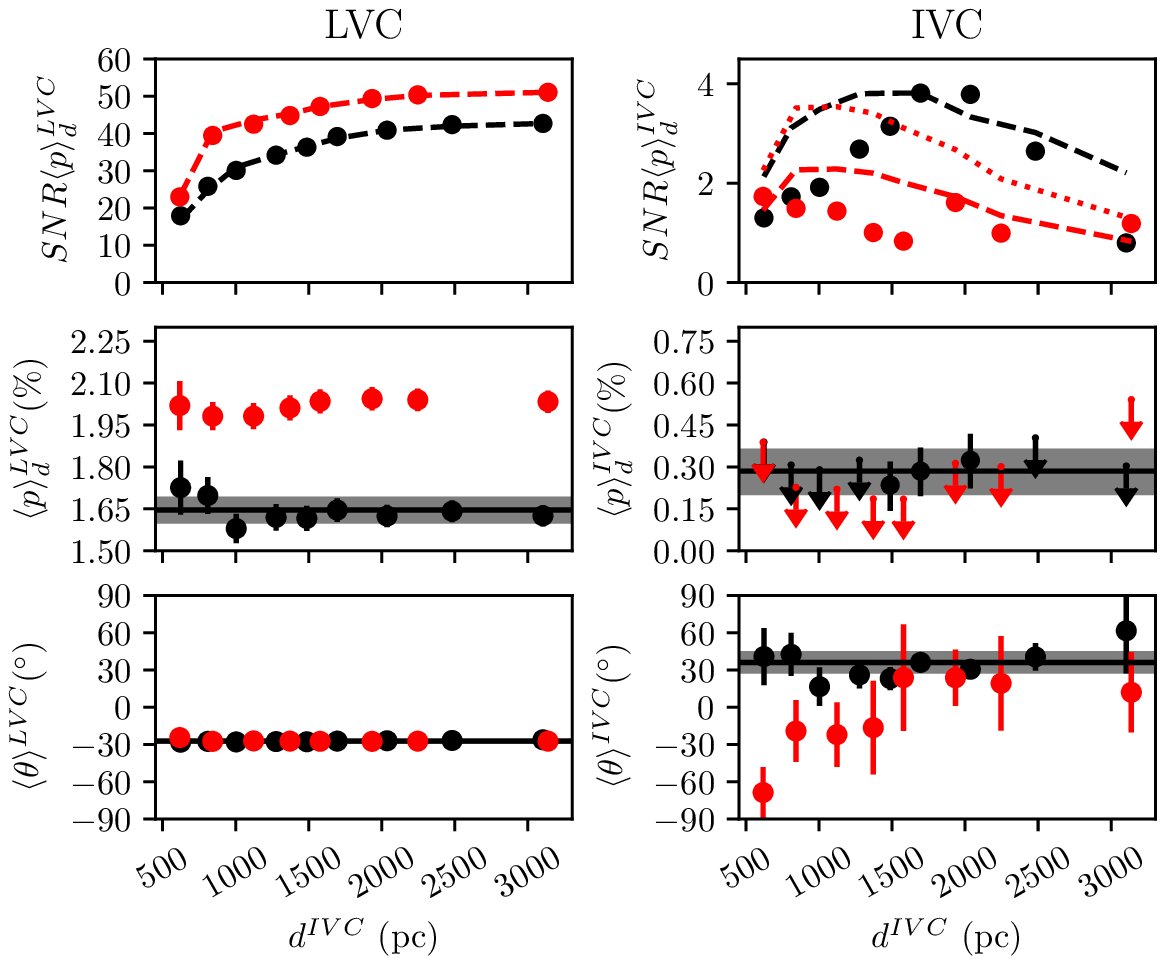}
\caption{Average polarization properties of the LVC (left) and IVC (right) after decomposition for different adopted distances to the IVC ($d^{IVC}$). From top to bottom: signal-to-noise ratio of $\langle p \rangle_d$, $\langle p \rangle_d$, $\langle \theta \rangle$. Black points refer to the 2-Cloud region while red refer to the 1-Cloud region. In the top panels, dashed and dotted lines show the effect of the uncertainty of $\langle p \rangle_d$ on the $SNR\langle p \rangle_d$, assuming a constant $\langle p \rangle_d$ as described in the text (red corresponds to the 1-Cloud region and black to the 2-Cloud region). In the middle panel, significant measurements of $\langle p \rangle_d$ are shown with their 1$\sigma$ uncertainties, while 2$\sigma$ upper limits are shown for measurements with $SNR\langle p \rangle_d < 3$. Black solid lines show the $\langle p \rangle_{d}$ (middle panel) and $\langle \theta \rangle$ (bottom panel) found in the 2-Cloud region for the $d^{IVC}$ where $SNR\langle p \rangle_d$ is maximum, while the gray bands mark the corresponding uncertainty. }
\label{fig:tomography_d}
\end{figure*}

Since there are two dominant clouds in the 2-Cloud region, there will be three populations of stars: foreground to both clouds (group 0), inter-cloud (group 1) and background to both clouds (group 2). The first are easy to distinguish from their negligible $p$, as we have seen in Section \ref{sec:dist}. With no exact distance to the second cloud, we cannot disentangle the two remaining populations. What we can do, is to assume a likely distance to the second cloud and calculate the decomposed mean polarization properties of each cloud under this assumption. In the following we shall evaluate how these properties depend on this assumption.

Let us assume a distance $d^{IVC}$ to the IVC. All stars with distances in the range [360 pc, $d^{IVC}$) are assigned to group 1. All that lie farther than $d^{IVC}$ are assigned to group 2. We find the weighted mean $q$ and $u$ in group 1 ($\langle q \rangle^{LVC}$, $\langle u \rangle^{LVC}$) and in group 2 ($\langle q \rangle^{IVC+LVC}$, $\langle u \rangle^{IVC+LVC}$). Then the mean $q$ and $u$ associated with the IVC only are:

\begin{eqnarray}
\langle q \rangle^{IVC}  &=& \langle q \rangle^{IVC+LVC} - \langle q \rangle^{LVC} \nonumber \\
\langle u \rangle^{IVC}  &=& \langle u \rangle^{IVC+LVC} - \langle u \rangle^{LVC}.
\label{eq:eqtomography}
\end{eqnarray}

These are used to calculate the mean polarization angle and fractional linear polarization (and associated uncertainties) of the LVC ($\langle \theta \rangle^{LVC}$, $\langle p \rangle^{LVC}$) and the IVC ($\langle \theta \rangle^{IVC}$, $\langle p \rangle^{IVC}$), from equations (\ref{eq:polarization_stokes}), (\ref{eq:angles_stokes}) and (\ref{eqn:sigmatheta}). 

The first assumed $d^{IVC}$ is set so that 10 stars are assigned to group 1 in order to obtain a statistically meaningful result for the mean polarizing properties of the LVC. Subsequent $d^{IVC}$ are assumed in steps of 10 stars. The results remain within the uncertainties if we instead select a step value of 5 or 15 (however, the uncertainties in the first case are larger). We perform the same analysis on the 1-Cloud sample.

The inferred properties of the mean polarization of each cloud for all assumed $d^{IVC}$ are shown in Figure \ref{fig:tomography_d}. The assumed distances of the IVC are in the range of 620 pc to 3.1 kpc. At distances less than 620 pc there are too few stars to be assigned to group 1, while at distances larger than 3.1 kpc there are too few stars to be assigned to group 2.

For all $d^{IVC}$, we find a highly significant (debiased) mean fractional linear polarization of the LVC, $\langle p \rangle_d^{LVC}$, (top left panel, Figure \ref{fig:tomography_d}). The signal-to-noise ratio of $\langle p \rangle_d^{LVC}$ ($SNR\langle p \rangle_d^{LVC}$) is higher than 18. This is the case for calculations done with 2-Cloud sample stars (black circles) and with 1-Cloud sample stars (red circles). The $\langle p \rangle_d^{LVC}$ remain constant for all $d^{IVC}$ (middle left panel). The same holds for the $\langle \theta \rangle^{LVC}$ (bottom left panel). Our choice of $d^{IVC}$ does not affect the mean polarization properties of the LVC. We note that the two regions differ in their $\langle p \rangle_d^{LVC}$. This is consistent with the fact that in the 1-Cloud region the LVC $N_{HI}$ is slightly higher than that in the 2-Cloud region (see also Section \ref{ssec:pebv}).

The increase of $SNR\langle p \rangle_d^{LVC}$ with $d^{IVC}$ (top left panel, Figure \ref{fig:tomography_d}) is caused by the reduction of the uncertainty on $\langle p \rangle_d^{LVC}$. At larger $d^{IVC}$, more stars are assigned to group 1, resulting in a reduced error on the ensemble average. We show that this is indeed the case in the top left panel of Figure \ref{fig:tomography_d}. We assume a constant value for $\langle p \rangle_d^{LVC}$ (equal to that found at the distance where the maximum $SNR\langle p \rangle_d^{LVC}$ is achieved: at $\sim 3100$ pc for both regions) and show the ratio of this value over the measured uncertainty of $\langle p \rangle_d^{LVC}$ at each $d^{IVC}$ (dashed lines). The measurements (circles) coincide with these lines, supporting our conclusion.

\begin{figure*}
\centering
\includegraphics[scale=0.9]{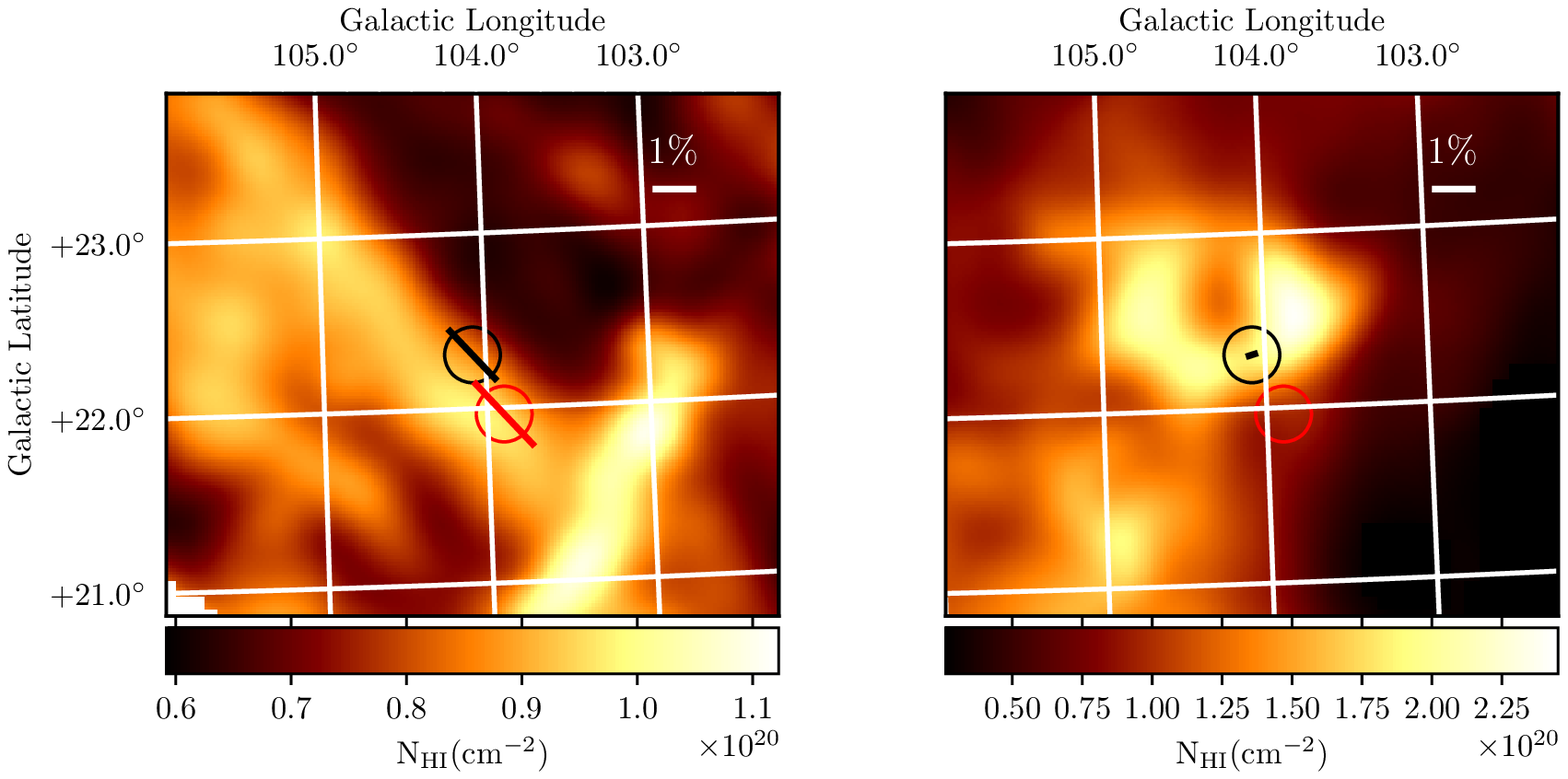}
\caption{A tomographic view of the mean orientation of the plane-of-the-sky magnetic field in each region. Line segments show the orientation of the field in the LVC (left, in both regions) and in the IVC (right, in the 2-Cloud region) and have length proportional to each cloud's $\langle p \rangle_d$. The values used are for the $d^{IVC}$ where max SNR$\langle p\rangle_d^{IVC}$ is achieved (see Figure \ref{fig:tomography_d}). Circles mark the 2-Cloud (black) and 1-Cloud (red) regions. The background images show $\rm N_{HI}$: (left) of the peak LVC emission (from integration within the range [-3.8,-1.2] km s$^{-1}$), and (right) of the IVC emission (within the range [-55,-41] km s$^{-1}$). A line of length 1\% is shown on the top right corner of each panel for scale.}
\label{fig:meanB}
\end{figure*}

The right panels of Figure \ref{fig:tomography_d} show the mean polarization properties inferred for the IVC using equations (\ref{eq:eqtomography}), (\ref{eq:polarization_stokes}) and (\ref{eq:angles_stokes}) for different assumed $d^{IVC}$. Here we find a significant difference between the two samples. From the 2-Cloud sample we find a signal-to-noise ratio of $\langle p \rangle_d^{IVC}$ ($SNR\langle p \rangle_d^{IVC}$) that depends strongly on $d^{IVC}$ (top right panel, black circles). At small $d^{IVC}$, the $\langle p \rangle_d^{IVC}$ is insignificant. As $d^{IVC}$ approaches $\sim$1.5 kpc, we find increasingly significant $\langle p \rangle_d^{IVC}$ (up to an $SNR\langle p \rangle_d^{IVC}$ of $\sim 4$). Then, at larger distances, the $SNR\langle p \rangle_d^{IVC}$ decreases. In contrast to this behaviour, the 1-Cloud sample does not yield any significant detection of $\langle p \rangle_d^{IVC}$ (top right panel, red circles). 

We investigate whether the observed behaviour of $SNR\langle p \rangle_d^{IVC}$ is a result of changes in the uncertainty of $\langle p \rangle_d^{IVC}$ as a function of assumed cloud distance. We set the value of $\langle p \rangle_d^{IVC}$ equal to that found at the $d^{IVC}$ where $SNR\langle p \rangle_d^{IVC}$ is maximum (for each region separately) and calculate the ratio of this value over the measured uncertainty of $\langle p \rangle_d^{IVC}$ at each $d^{IVC}$. The ratio is shown by the dashed lines in the top right panel of Figure \ref{fig:tomography_d}. By comparing the points in the 2-Cloud region (black circles) to the black dashed line it is clear that the observed variation of $SNR\langle p \rangle_d^{IVC}$ cannot be explained by a change in the uncertainties (which result from the distribution of stars along the line of sight). In particular, between 1 and 2 kpc the uncertainty remains approximately constant, while the $SNR\langle p \rangle_d^{IVC}$ increases significantly from 2 to 4. This would result from the presence of the IVC affecting the polarization of stars at these distances. 

In order to determine whether it is indeed the IVC that is causing the significant detection of $\langle p \rangle_d^{IVC}$, we look to the results in 1-Cloud region. Here, we do not detect significant $\langle p \rangle_d^{IVC}$ for any $d^{IVC}$. The dotted red line in the top right panel of Figure \ref{fig:tomography_d} shows the ratio of $\langle p \rangle_d^{IVC}$ found in the 2-Cloud region at the $d^{IVC}$ where $SNR\langle p \rangle_d^{IVC}$ is maximum ($\langle p \rangle_d^{IVC} = 0.29 \pm 0.08$(\%) at $d^{IVC} = 1695 $ pc) over the uncertainty on $\langle p \rangle^{IVC}$ for each $d_{IVC}$ in the 1-Cloud region. If the IVC were to induce $\langle p \rangle_d^{IVC}$ at the level found in the 2-Cloud region, we would expect to find a 3$\sigma$ detection within 1.5 kpc. This is not the case, as the observed $SNR\langle p \rangle_d^{IVC}$ are below 2 for all $d^{IVC}$. Since the IVC HI emission is significant in the 2-Cloud region but suppressed in the 1-Cloud region, we conclude that we have detected the signature of the IVC in the 2-Cloud region. We will show in Section \ref{ssec:distance} and Appendix \ref{sec:appendix} that this observed behavior of $SNR\langle p \rangle_d^{IVC}$ with assumed IVC distance is expected, and can help in determining the distance to the IVC.

The middle and bottom panels on the right (Figure \ref{fig:tomography_d}) show the $\langle p \rangle_d^{IVC}$ and $\langle \theta \rangle^{IVC}$ for different $d^{IVC}$. For the 2-Cloud region, both quantities are consistent within 1 $\sigma$ for all assumed distances to the IVC. The $\langle p \rangle_d^{IVC}$ is at the level of 0.29\% (for the $d_{IVC}$ where $SNR\langle p \rangle^{IVC}$ is maximal), a mere 18\% of that caused by the LVC in the 2-Cloud region (1.65\%). With such a difference in amplitude, it is not surprising that the effect of the IVC was not obvious when inspecting individual stellar polarizations with distance in Figure \ref{fig:ptdist}. Only upper limits on $\langle p \rangle^{IVC}$ can be placed in the 1-Cloud region.

The IVC differs not only in $p$ from the LVC, but also in $\theta$. With $\langle \theta \rangle^{IVC}$ 36$^\circ \pm$ 8$ ^\circ$ (for the $d^{IVC}$ where  $SNR\langle p \rangle_d^{IVC}$ is maximal), the IVC mean plane-of-the-sky magnetic field in the 2-Cloud region forms an angle of $\sim 60^\circ$ with that of the LVC (-27$^\circ \pm 1^\circ$).
Figure \ref{fig:meanB} shows the mean polarization properties of each cloud (after decomposition) on the plane of the sky. 
On the left, the line segments have length proportional to $\langle p \rangle_d^{LVC}$ (found in each region) and show the orientation of the mean (plane-of-the-sky) magnetic field of the LVC, as measured by $\theta^{LVC}$. We use the values for $\langle p \rangle_d^{LVC}$ and $\langle \theta \rangle^{LVC}$ found at the $d^{IVC}$ with the maximally significant detection of $\langle p \rangle_d^{IVC}$. The segment on the right shows the mean magnetic field orientation of the IVC ($\langle \theta \rangle_d^{IVC}$) and is on the same scale as the segments in the left panel. 

The $\langle \theta \rangle^{LVC}$ and $\langle \theta \rangle^{IVC}$ give the orientation of the mean plane-of-the-sky magnetic field in each cloud. This can be compared to the cloud morphology seen in $\rm N_{HI}$.  
The orientation of the HI emission of the IVC in the 2-Cloud region seems to follow the mean (plane-of-the-sky) magnetic field of the IVC. In the case of the LVC, we find the $\langle \theta \rangle^{LVC}$ in both regions to be aligned with the morphology of the emission within the velocity range where the HI spectrum peaks [-3.8,-1.2] km s$^{-1}$. The background image in the left panel of Figure \ref{fig:meanB} shows the $\rm N_{HI}$ from integrating within $\pm1$ velocity channel from the location of the $\rm T_b$ peak.

These findings are in agreement with the statistical alignment found between elongated structures in the diffuse ISM and the plane-of-the-sky magnetic field orientation \citep[with data covering a large sky fraction][]{clark,planckalignment}. Measures of the magnetic field orientation used in these works integrate along the line-of-sight (partially for starlight polarization and out to infinity for the polarization of dust emission). As a result, part of the observed spread in the relative orientation between matter and the magnetic field results from line-of-sight confusion. By applying a decomposition of the plane-of-the-sky magnetic field as a function of distance, as we have presented in this analysis, in a much larger sky fraction, this statistical correlation may become stronger. The alignment of the plane-of-the-sky magnetic field in the IVC with the curvature of the bubble-like gaseous structure resembles that found in works studying HII regions \citep[e.g.][]{chen}.

\begin{figure}
\centering
\includegraphics[scale=1]{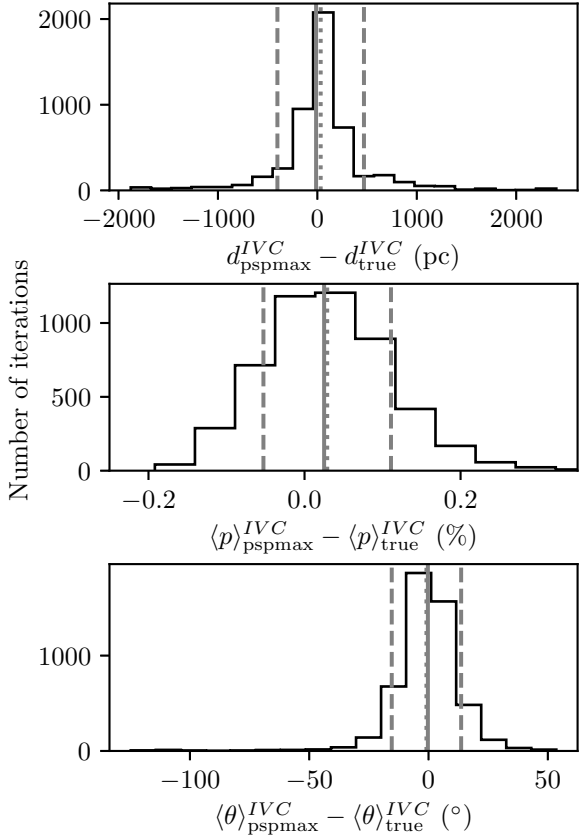}
\caption{Evaluation of the maximum $SNR\langle p\rangle_{IVC}$ as a predictor of the true properties of the IVC from 5000 realizations of the two-cloud model described in the text. Distributions of the difference between the predicted and true (top) distance to the cloud, (middle) mean fractional linear polarization, and (bottom) mean polarization angle. In all panels the solid gray line marks the median of the distribution, the dotted gray line marks the mean and the dashed lines bracket the range within $\pm$ 1 standard deviation.}
\label{fig:pspmaxdist}
\end{figure}

\subsection{Distance to the IVC from maximum $SNR\langle p\rangle_d^{IVC}$}
\label{ssec:distance}

In Section \ref{sec:tomography} we found that the $SNR\langle p\rangle_d^{IVC}$ varies with the assumed distance to the IVC. The gradual increase of $SNR \langle p\rangle_d^{IVC}$ in the 2-Cloud region as a function of $d^{IVC}$ and its subsequent decline (as well as the absence of this effect in the 1-Cloud region) makes it possible to constrain the distance to the IVC. 

In Appendix \ref{sec:appendix} we show analytically that the maximum $SNR\langle p\rangle^{IVC}$ occurs when the assumed distance to the IVC coincides with the true distance to the cloud (assuming a simplified distribution of measurement uncertainties). This can be understood intuitively, as at the true distance to the IVC the following two conditions are met: (a) the sample of stars used to determine the polarization properties of the IVC is free of contamination from sources that are foreground to the cloud (which are erroneously assigned to background sources, or group 2, at smaller assumed cloud distances), and (b) the maximum number of stars that are truly background to the cloud are used to calculate the ensemble average (at larger assumed cloud distances some stars that are in fact background to the IVC are erroneously assigned to the sample of foreground stars, or group 1).

In this section, we evaluate the $SNR\langle p\rangle_d^{IVC}$ as a distance indicator. To this end, we perform Monte Carlo simulations by creating mock observations of starlight polarization in the presence of two clouds with known distances. The first cloud is taken to lie at the distance of the LVC (360 pc). The second cloud is placed at distances in the range [700 pc, 2500 pc] in steps of 200 pc. The first and second cloud are taken to have mean $\langle p \rangle$ and $\langle \theta \rangle$ equal to those found for the LVC and IVC, respectively (Section \ref{sec:tomography}). We assume that $p$ and $\theta$ do not vary within the cloud, so that any variation will arise from measurement uncertainties. 

In each iteration of the model, we generate 103 measurements of starlight $p$ and $\theta$ (corresponding to the same sample size as in the 2-Cloud region) as follows. The stars are assigned the same distances and the same total uncertainty in $q$ and $u$ as in the observed sample. Each star $i$ that is background to the first cloud, but foreground to the second cloud is assigned a $q_i^{LVC}$ (and $u_i^{LVC}$)  drawn from a Gaussian distribution with mean equal to $\langle q \rangle^{LVC}$ ($\langle u \rangle^{LVC}$) and standard deviation equal to $\sigma_{q,i}$ ($\sigma_{u,i}$). Each star that is background to both clouds is assigned a total $q_i = q_i^{LVC} + q_i^{IVC}$ (and $u_i = u_i^{LVC} + u_i^{IVC}$). We draw $q_i^{LVC}$ ($u_i^{LVC}$) from a Gaussian distribution with mean equal to $\langle q \rangle^{LVC}$ ($\langle u \rangle^{LVC}$) and standard deviation equal to $\sigma_{q,i}/\sqrt{2}$ ($\sigma_{u,i}/\sqrt{2}$). The $q_i^{IVC}$ ($u_i^{IVC}$) are drawn from a Gaussian with mean $\langle q \rangle^{IVC}$ ($\langle u \rangle^{IVC}$) and standard deviation equal to $\sigma_{q,i}/\sqrt{2}$ ($\sigma_{u,i}/\sqrt{2}$). We select the standard deviation of the distribution so that the final uncertainty of this measurement ($\sqrt{\sigma^2_{q,i}/2 + \sigma^2_{q,i}/2}$) is equal to the observed $\sigma_{q,i}$ (and similarly for $\sigma_{u,i}$).  

Then, we follow the process outlined in Section \ref{sec:tomography}: we assume different distances to the second cloud (in distance steps of 10 stars), assign stars to two groups, and compute the ensemble average $\langle q \rangle$ and $\langle u \rangle$ of each group. Finally, we find the mean polarization properties of the first and second cloud (decomposed along the line of sight). For each iteration, we find the assumed distance to the second cloud, $d_{\rm pspmax}^{IVC}$, where the $SNR\langle p \rangle_d^{IVC}$ of the mock dataset is maximum, as well as the fractional linear polarization and polarization angle of the second cloud at that assumed distance ($\langle p \rangle_{\rm pspmax}^{IVC}$ and $\langle \theta \rangle_{\rm pspmax}^{IVC}$).

We compare these quantities with the true properties of the second cloud ($d_{\rm true}^{IVC}$, $\langle p \rangle_{\rm true}^{IVC}$, $\langle \theta \rangle_{\rm true}^{IVC}$) in Figure \ref{fig:pspmaxdist}, which shows results from 5000 iterations of the model. The distance where $SNR\langle p \rangle_d^{IVC}$ is maximum is a good indicator of the true distance for our simulations (top panel, Figure \ref{fig:pspmaxdist}). The distribution of $d_{\rm pspmax}^{IVC} - d_{\rm true}^{IVC}$ has a mean of 33 pc, a median of -13 pc and a standard deviation of 440 pc. The standard deviation is slightly larger than the typical sampling of $\sim 200-300$ pc in $d^{IVC}$ (corresponding to a step of 10 stars in our sample).

The average polarization properties of the second cloud are accurately recovered at the assumed distance $d_{pspmax}^{IVC}$ (middle and bottom panels, Figure \ref{fig:pspmaxdist}). The standard deviation of the distribution of $\langle p\rangle_{pspmax}^{IVC} - \langle p\rangle_{\rm true}^{IVC}$ is comparable to the uncertainty of the observed $\langle p \rangle^{IVC}$ (0.082\% compared to 0.075\%). In the case of the distribution of $\langle \theta\rangle_{\rm{pspmax}}^{IVC} - \langle \theta\rangle_{\rm{true}}^{IVC}$, the standard deviation is twice as much as the uncertainty of $\langle \theta \rangle^{IVC}$ (16$^\circ$ compared to 8$^\circ$).

The spread of the distribution of $d_{\rm pspmax}^{IVC} - d_{\rm true}^{IVC}$ can be used as an estimate of the accuracy of the method in determining the true distance to the cloud. This spread depends slightly on the choice of distance sampling. When performing the tomographic decomposition, we assumed cloud distances with a step of 10 stars. If we change this value to 30 stars, the standard deviation of the distribution of $d_{\rm pspmax}^{IVC} - d_{\rm true}^{IVC}$ increases by 15\%, as one would expect due to the coarser sampling. The median of the distribution shifts by 100 pc (from -13 to -96 pc), while the mean changes from 33 pc to -190 pc (within the 1 $\sigma$ of 510 pc). The median and standard deviation of the distribution of $\langle p \rangle_{\rm pspmax}^{IVC} - \langle p \rangle_{\rm true}^{IVC}$ and $\langle \theta \rangle_{\rm pspmax}^{IVC} - \langle \theta \rangle_{\rm true}^{IVC}$ vary by less than 15$\%$.

The accuracy of $d_{\rm pspmax}^{IVC}$ as an indicator of the true distance to the cloud depends on $d_{\rm true}^{IVC}$. At small $d_{\rm true}^{IVC}$, the distribution of $d_{\rm pspmax}^{IVC} - d_{\rm true}^{IVC}$ is asymmetric with a long tail towards larger values. The opposite happens at large $d_{\rm true}^{IVC}$ (a tail develops towards smaller values). This is most likely due to the distribution of stellar distances in our sample, which peaks at $\sim 1$ kpc. To evaluate the accuracy of this method in situations with different cloud properties, and different stellar distance distributions, further work is needed. 

The tests presented here show that, for the specific case of the observed 2-Cloud region, the $d_{\rm pspmax}^{IVC}$ can be used to constrain the true distance to the IVC. Since the maximum $SNR\langle p \rangle^{IVC}$ is found at $\sim 1700$ pc, and our tests show a typical uncertainty (standard deviation of $d_{\rm pspmax}^{IVC} - d_{\rm true}^{IVC}$) of 440 pc, we conclude that the IVC is most likely located within the range $\sim$[1250 - 2140] pc.

\section{Discussion} \label{sec:discussion}

\begin{table*}
    \centering          
    \caption{Properties of the clouds in the 2-Cloud and 1-Cloud regions\footnote{E(B-V)$\rm_{HI}$ is the reddening derived from $\rm N_{HI}$, and E(B-V)$\rm _d$ is the total reddening for the specified component, and $\rm E(B-V)_{d,los}$ is the total reddening of the sightline. For details see appendix \ref{sec:appendixB}.}}             
    \begin{tabular}{l c c c c c c}   
      \hline\hline                     
       Region (vel. component) & Velocity range & $\rm N_{HI} \times 10^{20} $&  $\rm E(B-V)_{HI}$ & $\rm E(B-V)_d$& $\rm \frac{E(B-V)_{HI}}{E(B-V)_{d,los}}$ & $f_{mol}$\\ 
                               & (km $\rm{s^{-1}}$)                & $\rm (cm^{-2})$  & (mag) within 2mmag & (mag) $\pm$0.01 &  &\\
       \hline  
       2-Cloud (IVC) 		   & [-55, -41] 	    & 1.8	& 0.02  &	$\geq 0.02$ & 0.10 & $\geq 0$\\
       2-Cloud (LVC) 		   & [-12, 5] 	    & 3.5	& 0.04  &	$\leq 0.16$ & 0.19 &	 $\leq 0.75$\\
       2-Cloud (entire los) &  [-600, 600]  	& 8.2	& 0.09  &	0.21 & 0.45	& $-$\\
       1-Cloud (IVC) 	   & [-55, -41] 	    & 0.9   	& 0.01 &    $ 0.01$ & 0.05 & $\sim 0$  \\
       1-Cloud (LVC) 	   & [-12, 5]  	    & 4.0	& 0.05 & 	$ 0.18$ & 0.20	& $0.75$ \\
       1-Cloud (entire los) & [-600, 600]    & 7.8   & 0.09 &    0.23 & 0.40  & $-$\\
       \hline          
    \end{tabular} 
    \label{tab:NH}
  \end{table*} 
  
\subsection{Mean polarizing efficiency of the two clouds}
\label{ssec:pebv}

Measurements of $p$ for individual stars are bounded by an upper envelope with respect to reddening, E(B-V) \citep{hiltner}:
\begin{equation}
p_{max} = \rm 9 E(B-V) (\%/mag),
\label{eqn:pebv}
\end{equation}
which describes the maximum polarizing efficiency of the ISM per unit dust column.
Recently, \citet{Skalidis2018} presented evidence that this upper envelope differs at very low extinction. The upper envelope was revised by \citet{planckxii} using the polarization measurements of \citet{berdyugin} for stars within 600 pc at high-Galactic latitude and \textit{Planck}\footnote{\textit{Planck} (http://www.esa.int/Planck) is a project of the European Space Agency (ESA) with instruments provided by two scientific consortia funded by ESA member states and led by Principal Investigators from France and Italy, telescope reflectors provided through a collaboration between ESA and a scientific consortium led and funded by Denmark, and additional contributions from NASA (USA).} sub-mm polarization. They propose $p_{max} = \rm 13 E(B-V) (\%/mag)$. 

With our tomographic decomposition of the polarization properties of the IVC and LVC, we can compare the effectiveness of these two \textit{individual} clouds in polarizing starlight to the aforementioned line-of-sight-integrated relations. We will therefore compare the average $p$ found in section \ref{sec:tomography} \textsl{for each cloud} to the average reddening \textsl{of each cloud} in the observed regions. For this purpose, we must obtain estimates of the mean reddening of each cloud, which is straightforward for the 1-Cloud region, but not as simple for the 2-Cloud region. In the following, we describe how we obtain our estimates of the per-cloud reddening in each region.

In the diffuse ISM, reddening is well correlated with the hydrogen column density, $\rm N_{HI}$ \citep[e.g.][]{Bohlin}. We can therefore use the HI emission data from the HI4PI survey (Section \ref{sec:HIdust}), to obtain an estimate of the reddening caused by each cloud separately. By integrating the HI emission over the range of velocities of the IVC, and the LVC, we find $\rm N_{HI}$ of the order of $\sim 10^{20} \rm cm^{-2}$ for both clouds. This column density is where the transition from atomic to molecular hydrogen is found to occur \citep[e.g.][]{gillmon}. For this reason, deriving an estimate of the reddening solely from $\rm N_{HI}$ may bias the result to lower reddenings. We take into account auxiliary information from FIR dust emission from \textit{Planck} and derive limits on the reddening of each cloud in Appendix \ref{sec:appendixB}. The properties of the two clouds are listed in Table \ref{tab:NH}. 

\begin{figure}
\centering
\includegraphics[scale=1]{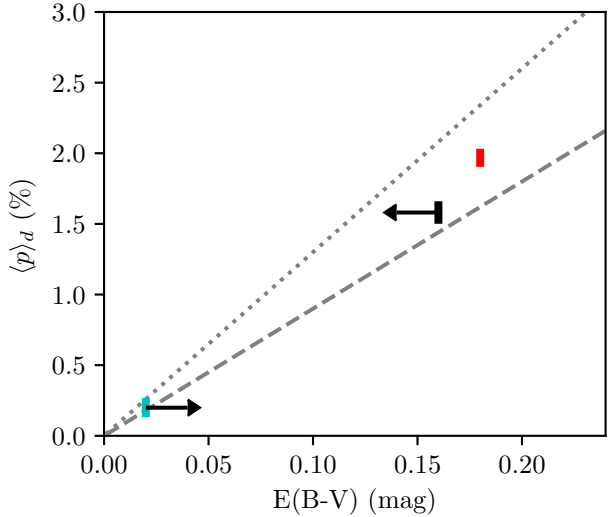}
\caption{The mean polarizing efficiency of each cloud, inferred from the relation between the debiased average fractional linear polarization $\left\langle p\right\rangle_d$ (from Section \ref{sec:tomography}) versus the average reddening (values on the horizontal axis refer to the \textsl{total} reddening obtained from the analysis in Appendix \ref{sec:appendixB} and shown in the $\rm E(B-V)_d$ column of Table \ref{tab:NH}). The extent of the symbols covers the range of possible values. Light blue: IVC in the 2-Cloud region. Red: LVC in the 1-Cloud region. Black: LVC in the 2-Cloud region. The relations $p_{max} = \rm 9 E(B-V)$ and $p_{max} = \rm 13 E(B-V)$ are shown with dashed and dotted lines, respectively.}
\label{fig:pebv}
\end{figure}

The unknown distance to the IVC introduces an uncertainty in the value of $\left\langle p\right\rangle$ for the LVC and IVC (see Fig. \ref{fig:tomography_d}). We take this into account as follows: We only use the statistically significant values shown in Figure \ref{fig:tomography_d} ($\left\langle p\right\rangle/\sigma_{\left\langle p\right\rangle_d} \geq 3$). 
We show the range of values from $\min\left\lbrace\left\langle p\right\rangle_d\right\rbrace - \sigma_{\left\langle p\right\rangle}$ to $\max\left\lbrace\left\langle p\right\rangle_d\right\rbrace + \sigma_{\left\langle p\right\rangle}$. In Figure \ref{fig:pebv}, the light blue rectangle covers the range in $\left\langle p\right\rangle_d$ of the IVC. The range of values for the LVC is shown in black for the 2-Cloud region and in red for the 1-Cloud region. 

\begin{table*}
\centering
\caption{Comparison between properties of polarized thermal dust emission measured by \textit{Planck} at 353 GHz and starlight polarization.}
\begin{tabular}{|c|c|c|c|c|c|}
\hline
Region & $p_{353}$ & $SNRp_{353}$ & $\chi_{\rm gal,353}+90^\circ$ & mean $\theta_{\rm gal,far}$ & Cloud $\left\langle\theta\right\rangle_{\rm gal}$\footnote{All angles are measured with respect to the Galactic reference frame (IAU convention), specified by the subscript `gal'.} \\
\hline
\hline
2-Cloud & 10\% & 19 & 45$^\circ$ $\pm$1.6$^\circ$ & 47$^\circ \pm 1^\circ$ & 42$^\circ$ $-$ 44$^\circ$ (LVC)  \\ 
        &      &     &                         &                         & 87$^\circ$ $-$ 132$^\circ$ (IVC)\\
1-Cloud & 10\% & 23 & 43$^\circ$ $\pm$1.3$^\circ$ & 45$^\circ \pm 1^\circ$  & 42$^\circ$ $-$ 45$^\circ$ (LVC)\\ 
\hline
\end{tabular}
\label{tab:planckcompare}
\end{table*}

From Fig. \ref{fig:pebv}, we find that both clouds seem to be highly efficient in polarizing starlight as they fall between the original upper envelope of \citet{hiltner} and the revised one from \cite{planckxii} dashed and dotted lines in Fig. \ref{fig:pebv}, respectively). Our measurements refer to the \textit{mean} polarization induced by each cloud. 
Therefore, each cloud is potentially capable of inducing $p$ higher (and lower) than $p_{max}$. The mean polarizing efficiencies of the two clouds depend on the molecular content of the IVC. If the IVC has nonzero molecular content, its reddening will move towards higher values while the LVC reddening will be pushed to lower values (and thus the LVC will be pushed to higher polarizing efficiencies). 

The fact that both clouds are very efficient in polarizing starlight can help us constrain some of the physical properties of the clouds. As shown by \cite{LeeDraine}, ISM-induced $p$ follows the relation:
\begin{equation}
p = p_{max} R \frac{3}{2}(\left\langle \cos^2\delta\theta\right\rangle-\frac{1}{3}) \cos^2\gamma , 
\label{eqn:leedraine}
\end{equation}
where $p_{max}$ reflects the polarizing capability of the dust grains due to their geometric and chemical characteristics, $R$ quantifies the degree of alignment of the grains with the magnetic field, $\gamma$ is the inclination angle between the magnetic field and the plane of the sky, and $\delta\theta$ is the angle between the direction of the field at any point along the line of sight and the mean field direction. The angular brackets denote an average along the line of sight. 

The high mean $p$ of the IVC and LVC therefore suggests that in both clouds the depolarizing factors are minimal. First, the 3D magnetic field orientation must lie close to the plane of the sky ($\gamma \sim 0$). Second, any tangling of the field (variation of the orientation along the los) must be small. The ordered component of the field within each cloud must dominate over the random component (otherwise fluctuations in the orientation would be significant). 
This agrees with our finding that the magnetic field, as projected on the plane of the sky, is ordered: the distribution of polarization angles is narrow Figure \ref{fig:ITqu}. Since the LVC is dominating the signal, the ordered polarization segments seen in Figure \ref{fig:segs} reflect the strength of the magnetic field in this cloud. 

\subsection{Comparison to polarized thermal dust emission from \textit{Planck}}
\label{subsec:planck}

Since the polarization of starlight in absorption is connected to the polarized thermal emission from dust in the ISM, we wish to compare our measurements of the mean optical polarization in the observed regions to those of the emission from the \textit{Planck} mission at 353 GHz. 

We use the Planck-HFI full mission data at 353 GHz \citep{planckhfi}, which have a native resolution (beam FWHM) of 4.8$\arcmin$ and are sampled on a HEALPix grid with NSIDE 2048. At this native resolution, the Planck uncertainties are high. 
To increase the SNR, we smooth each map using the SMOOTHING utility of the healpy python library, which performs smoothing in spherical harmonic space. The final angular resolution of the maps is 15$\arcmin$ (FWHM). We downgrade these smoothed maps from the native NSIDE 2048 to NSIDE 512, resulting in a final pixel angular size of 6.6$\arcmin$. When smoothing the maps we have taken into account the rotation of the celestial reference frame in each pixel, as discussed in appendix A of \citet{planckxix}. The effect on the smoothed values of the Stokes parameters is minimal, as the field under examination is far from the Galactic poles (b = 22$^\circ$) and small in angular extent.

In each pixel we find the polarized intensity: $P=\sqrt{Q^2+U^2}$, and the polarization angle with respect to the North Galactic Pole (NGP): $\chi_{gal} = 0.5\arctan(-U/Q)$ (IAU convention), where we use the two-argument arctangent that lifts the $\pi$ ambiguity. The uncertainties on these quantities are found using equations (B5) and (B4) of \citep{planckxix}. Finally, we construct a map of $p_{353} = P/I_{353}$, where $I_{353}$ is the total intensity at 353 GHz. A large uncertainty on $p_{353}$ comes from the zero-point offset of $I_{353}$ \citep[][]{planckxii}. We do not follow any of the suggested corrections as we do not use the value of $p_{353}$ in what follows.

We report the (weighted) mean value of each quantity within the 2-Cloud and 1-Cloud regions in Table \ref{tab:planckcompare}. We compare with values obtained from starlight polarization by converting $\theta$ to angle with respect to the NGP, $\theta_{gal}$, following \cite{erratum}. As the \textit{Planck} data are integrated along the line of sight, to make a fair comparison with the optical polarization data we average over the furthest stars in each region (with distances $>2$ kpc). The Table columns are: (1) fractional linear polarization of thermal dust emission ($p_{353} = P_{353}/I_{353}$) and (2) its signal-to-noise ratio ($SNRp_{353} = p_{353}/\sigma_{p,353}$); (3) polarization angle of dust emission, $\chi_{\rm gal, 353}$ (rotated by 90$^\circ$); (4) starlight polarization angle averaged over stars farther than 2 kpc, $\theta_{\rm gal, far}$; (5) most likely mean polarization angle of each cloud ($\left\langle\theta\right\rangle_{\rm gal}$) from section \ref{sec:tomography}. 

The polarization angle found by \textit{Planck} remains constant (within the uncertainties) between the two regions. By rotating by 90$^\circ$ to compare with the starlight polarization data, we find that the \textit{Planck} polarization angle is in agreement with the mean $\theta_{gal}$ found in the two regions. It is also consistent with the mean value found for the LVC. This is not surprising, as the LVC dust column is as much as twice that of the IVC (Table \ref{tab:NH}) and was found to dominate the signal in starlight polarization. 

\subsection{Frequency dependence of the dust emission polarization angle}
\label{ssec:decorrelation}

As the dust emission provides line-of-sight integrated information, there is no way of detecting the presence of the two distinct clouds at a single frequency. However, the existence of these two strikingly different sources of polarized signal could manifest itself as a variation of the \textsl{polarization angle as a function of frequency}. This effect, which has a well-known counterpart in the optical\footnote{As discussed by e.g. \cite{martin}, circular polarization (or a wavelength dependence of the linear polarization angle) can arise from the passage of light through two media with both (a) different dust properties, parametrized by the wavelength at which maximum $p$ occurs \citep[$\lambda_{max}$][]{serkowski} and (b) different grain alignment (magnetic field) orientations.} \citep[e.g.][]{serkowski1962,treanor1963,coyne,martin}, has been pointed out by \citet{tassis} and discussed in the context of CMB-foreground subtraction by various works \citep[e.g.][]{poh,hensley,planckxi2018}. For a significant difference between the polarization angle at different frequencies to occur, two conditions must be met: the magnetic fields of the two clouds must have significantly different orientations projected on the plane of the sky and the dust temperatures and/or dust emission spectral indices of the two clouds must not be identical. The first condition is met in our selected 2-Cloud region: we have found a difference of $\sim 60^\circ$ between the IVC and LVC polarization angle. 

\begin{figure*}
\centering
\includegraphics[scale=1]{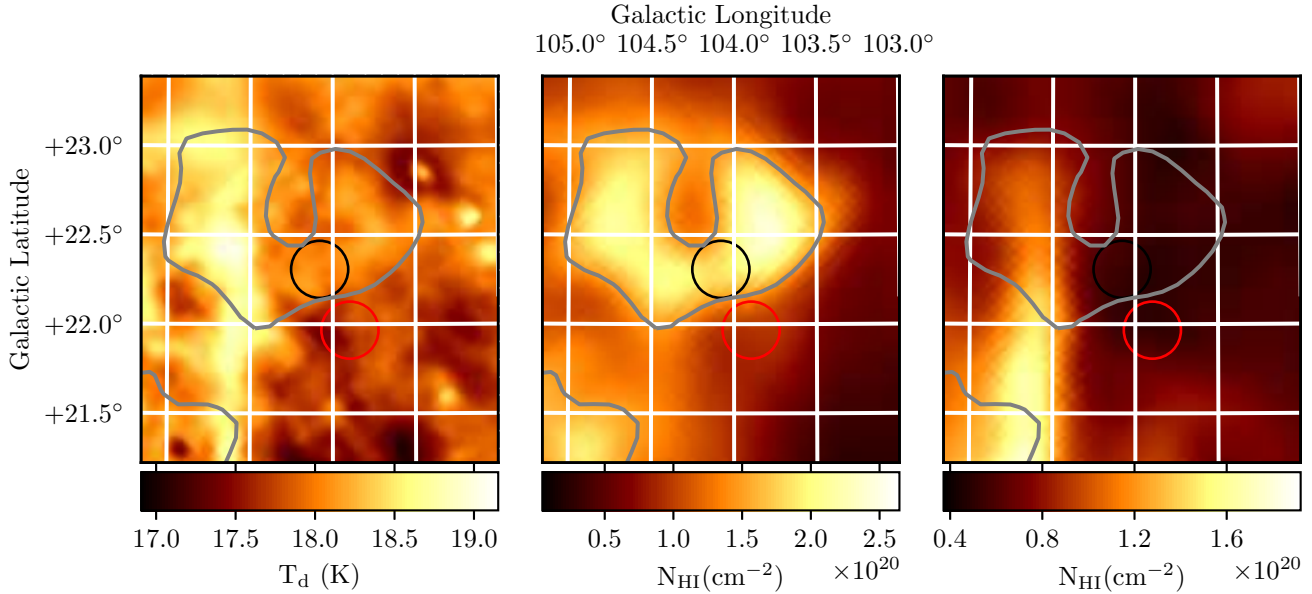}
\caption{Evidence for differences in dust temperature along the line of sight. Left: Dust temperature from \citet{planckxlviii}. Middle: Column density of HI from the HI4PI data, for the range of velocities of the IVC [-55, -41] km $\rm{s^{-1}}$. Right: Column density of HI, from the HI4PI data for the range of velocities [-30,-20) km $\rm{s^{-1}}$, where a third component of emission is present in part of the map. The gray contour outlines the emission from the IVC while the black and red circles mark the two regions observed in this work.}
\label{fig:planck}
\end{figure*}

We can investigate whether the second condition is also met by invoking supplementary information. To this end, we use the map of dust temperature, $\rm T_d$, presented in \citet{planckxlviii}. This map was derived by fitting a modified black body (MBB) to each pixel of the component-separated multi-frequency maps of dust emission (obtained through the Generalized Needlet Internal Linear Combination $-$GNILC$-$ method). This is the highest resolution map of $\rm T_d$ and is free of contamination from cosmic infrared background anisotropies \citet{planckxlviii}.

Figure \ref{fig:planck} (left) presents $\rm T_d$ within 1 degree centred on the 2-Cloud region. The middle panel shows $\rm N_{HI}$ derived from integrating the emission in the area within the velocity range [-55, -41] km $\rm{s^{-1}}$ (where the IVC dominates). The two maps show some degree of spatial correlation (Pearson correlation coefficient = 0.45). The IVC is outlined by the central gray contour. With further inspection one can notice the outline of the IVC towards the center and right of the field also in the $\rm T_d$ image. 

The $\rm T_d$ map shows a prominent feature running vertically throughout the left portion of the area, outside the two regions observed in this work. We have searched for a counterpart in the HI emission, and have found one within the range of velocities [-30, -20) km $\rm{s^{-1}}$. We show the $\rm N_{HI}$ map of emission integrated in this range in the right panel of Figure \ref{fig:planck}. 

Both the IVC and this last component seem to have influenced the single-MBB fit towards yielding higher temperatures. We note that the LVC covers the entire area shown in this map. This is a strong indication that the two clouds appearing in the $\rm N_{HI}$ maps of Figure \ref{fig:planck} have $\rm T_d$ that is higher from that of the local emission. One possible interpretation is the existence of abundant molecular material in the LVC (in contrast to the IVC), which must produce stronger shielding from the interstellar radiation field compared to the IVC. The analysis by \citet{planckdusthalo} for a large sample of IVCs also found these clouds to have higher $\rm T_d$ than local clouds. 

Since the IVC is significantly subdominant compared to the LVC in our selected sightline, the effect of rotation of the polarization angle with frequency may be difficult to detect. A more promising case may be that of the prominent feature at velocities [-30,20) km $\rm{s^{-1}}$, as its effect on the MBB fit is more pronounced than that of the IVC. This cloud, however, lies outside the areas where we have measured starlight polarization in this work.

Having found evidence that the temperature of the IVC differs from that of the LVC, we proceed to estimate the frequency dependence of the dust emission polarization angle. We model the total intensity and polarized emission of each cloud and derive an expression for the polarization angle as a function of frequency, $\nu$, and cloud parameters (namely, the dust temperature and spectral index in each cloud, ${\rm T^{C_1}_d, T^{C_2}_d}, \beta^{C_1}, \beta^{C_2}$, and the ratio of polarized intensities of the two clouds $r_\nu = P^{C_1}_\nu/P^{C_2}_\nu$) in Appendix \ref{sec:appendixC}. 

We shall examine the frequency dependence of the polarization simply by estimating the difference between the polarization angle at two frequencies: 70 GHz and 353 GHz. The former frequency is relevant for CMB-foreground subtraction, as it coincides with the minimum contribution of foregrounds to the polarized signal of the CMB, as modelled by \citet{planckx}. The latter frequency is that for which the properties of the polarized dust emission are best understood through measurements from \textit{Planck} \citep{planckxii}. From equation \ref{eqn:chinur}, the difference between the polarization angle of emission between these two frequencies is:
\begin{eqnarray}
\chi_{353} - \chi_{70} &=& \frac{1}{2} \arctan{ \frac{r_{353} \sin{2\chi^{IVC}} + \sin{2\chi^{LVC}} }{ r_{353} \cos{2\chi^{IVC}} + \cos{2\chi^{LVC}} }} \nonumber \\
& & - \frac{1}{2} \arctan{ \frac{r_{70} \sin{2\chi^{IVC}} + \sin{2\chi^{LVC}} }{ r_{70} \cos{2\chi^{IVC}} + \cos{2\chi^{LVC}} }}.
\label{eqn:deltachi}
\end{eqnarray}
We use the symbol $\chi$ to specify the polarization angle of the dust emission, and reserve $\theta$ for the polarization angle measured in the optical. The polarization angle of the dust emission from a single cloud (e.g. the LVC) is related to the polarization angle in the optical (due to the presence of the same cloud) by: $\chi^{LVC} = 90^\circ + \theta^{LVC}$, and similarly for the IVC (see appendix \ref{sec:appendixC}).

We have measured the fractional linear polarization $p$ (in the optical) in each cloud (Section \ref{sec:tomography}), which can be used to infer $r_{353}$ by means of the ratio of polarized intensity $P_{353}$ at 353 GHz over $p$ measured in the optical, considering the following. Recently, \cite{planckxii} showed that starlight fractional linear polarization in the V band ($p_V$) and $P_{353}$ towards thousands of diffuse sightlines are well correlated: $P_{353}/p_V = 5.38 \pm 0.03 $ MJy/sr. Thus, when considering ensembles of sightlines, starlight $p_V$ can be used to predict $P_{353}$. This relation cannot, however, be taken to hold exactly when studying individual clouds (two in the case of our selected region). Even in the line-of-sight-integrated data used in \cite{planckxii}, deviations from the relation can be seen. There are two factors that can be causing the observed scatter: (a) a star may not be tracing the entire sightline and (b) a specific sightline may contain different dust properties (e.g. temperature, spectral index) than the sky-averaged values. Factor (a) is most likely subdominant for the sample used in their work, as stars were shown to trace $\sim$80\% of the column of the sightline. Another piece of evidence that supports the view that $P_{353}/p_V$ depends on the specific characteristics of a cloud is that it is shown to vary as a function of column density \citep[for low column densities, Figure 27 of][]{planckxii}. 

\begin{table}
\centering
\caption{Values of $r_{353}$ found by assuming different relations between $P^{cloud}_{353}$ and $\left\langle p\right\rangle^{cloud}_R$.}
\begin{tabular}{|c|c|}
\hline
$P^{cloud}_{353}$ (MJy sr$^{-1}$) & $r_{353}$ \\
\hline
\hline
$P^{LVC} = 5.38 \left\langle p\right\rangle^{LVC}_R   $  &    \\ 
$P^{IVC} = 5.38 \left\langle p\right\rangle^{IVC}_R   $  &  0.17      \\
\hline
$P^{LVC} = 5.38\left\langle p\right\rangle^{LVC}_R - 0.015$  &    \\ 
$P^{IVC} = 5.38\left\langle p\right\rangle^{IVC}_R - 0.015$  &  0.001      \\
\hline
$P^{LVC} = 5.38\left\langle p\right\rangle^{LVC}_R + 0.015$  & \\ 
$P^{IVC} = 5.38\left\langle p\right\rangle^{IVC}_R - 0.015$  & 0.001  \\
\hline
$P^{LVC} = 5.38\left\langle p\right\rangle^{LVC}_R + 0.015$ &  \\
$P^{IVC} = 5.38\left\langle p\right\rangle^{IVC}_R + 0.015$ & 0.29      \\
\hline
$P^{LVC} = 5.38\left\langle p\right\rangle^{LVC}_R - 0.015$ & \\
$P^{IVC} = 5.38\left\langle p\right\rangle^{IVC}_R + 0.015$  & 0.41\\
\hline
\end{tabular}
\label{tab:r353}
\end{table}

To take the aforementioned into account, we will assume the following two scenaria: 
\begin{itemize}
\item[Case A]: The ratio of $P_{353}/p_V$ is the same in both clouds and equal to the mean value measured by \cite{planckxii} for diffuse sightlines: $P^{C_1}_{353}/p^{C_1}_V = P^{C_2}_{353}/p^{C_2}_V = 5.38 \pm 0.03 $ MJy/sr
\item[Case B]: The ratio of $P_{353}/p_V$ is different in each cloud. $P_{353}$ can take values within the range observed for a given $p_V$ 
\citep[5.38$ p_V\pm$0.015 MJy/sr from Figure 27 of][]{planckxii}. 
\end{itemize}
Since starlight polarization measured in the $R$ and $V$ bands varies within 10\% \citep{serkowski}, we will take $p_R = p_V $ for simplicity. We use the subscript $R$ from now on to refer to optical measurements and distinguish from the ratio of polarized intensity over total intensity in emission at frequency $\nu$ ($p_\nu$). We assume that the ratio $P_{353}/p_R$ does not vary within a single cloud, so that the mean $\left\langle p\right\rangle^{cloud}_R$ (measured in Section \ref{sec:tomography}) can be used to predict the mean $P^{cloud}_{353}$. 

\begin{figure*}
\centering
\includegraphics[scale=1]{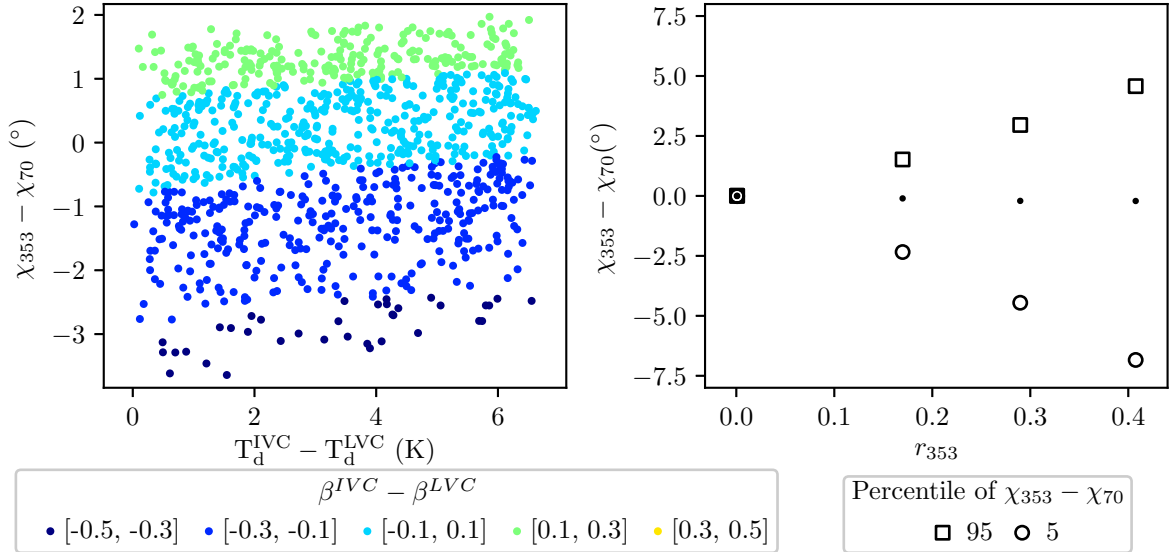}
\caption{Left: Difference between predicted polarization angles of dust emission measured at 353 GHz and at 70 GHz, as a function of the temperature difference between the two clouds. Values are from 1000 realizations of the model described in Section \ref{subsec:planck} (CaseA). Colours correspond to different values of the difference of cloud spectral indices $\beta_{IVC}-\beta_{LVC}$. Right: 5 (circles) and 95 (squares) percentiles of the distribution of angle differences between 353 GHz and 70 GHz for different values of $r_{353}$, arising from different assumptions for the ratio of polarized intensity in emission over $p$ in the optical (Case B). The filled dots show the mean. The case shown on the left panel corresponds to $r_{353} = 0.17$.}
\label{fig:decorrelation}
\end{figure*}

For case A: $P^{LVC}/\left\langle p\right\rangle^{LVC}_R = P^{IVC}/\left\langle p\right\rangle^{IVC}_R$, and equation \ref{eqn:ratioP} gives $r_{353} = \left\langle p\right\rangle^{IVC}_R  / \left\langle p\right\rangle^{LVC}_R  = 0.28/1.65 = 0.17$. For case B, we take all four combinations of the extreme cases of $P^{cloud}{353} = 5.38 \left\langle p\right\rangle^{cloud}_R \pm 0.015$ MJy/sr. The values of $r_{353}$ for all cases are summarized in Table \ref{tab:r353}.

To evaluate the model, we must assume values for the dust temperature and spectral index in each cloud. The dust temperature in the LVC is taken to lie within the range of observed $\rm T_d$ in the 1-Cloud region $\rm T^{LVC}_{d,min} = 17.3 \, K, T^{LVC}_{d,max} = 17.9$ K. The IVC must have a higher dust temperature, so $\rm T^{IVC}_{d,min} = T^{LVC}_{d,max}$. A reasonable upper limit on the dust temperature for the IVC can be taken from the studies of IVCs in \citet{planckdusthalo, planckxvii}: $\rm T^{IVC}_{d,max} = 24$ K. The LVC spectral index can be constrained by use of the maps published by \citet{planckx} and \citet{planckxlviii}. In the first map, which was constructed using the COMMANDER component separation method, we find a range of values of $\beta$ in the 1-Cloud region: $\beta^{LVC}_{min}, \beta^{LVC}_{max} = 1.49, 1.58$, and a mean value of 1.53. In the second map, constructed using the GNILC method, we find $\beta^{LVC}_{min}, \beta^{LVC}_{max} = 1.63, 1.68$, and a mean of 1.64. We thus constrain the $\beta^{LVC}$ to be in the range [1.49, 1.68]. For the IVC, we can use values of the spectral index from \citet{planckxvii}. We take $\beta^{IVC}$ to be within two standard deviations from the mean found in \citet{planckxvii}, i.e. $\beta_{min}, \beta_{max} = 1.3, 1.8$. 

We calculate equation \ref{eqn:deltachi} $10^3$ times, for $r_{353} = 0.17$ (i.e. for Case A, where both clouds are taken to have the same ratio $P_{353}/p_R$. In each realization, the temperature and spectral index of each cloud are drawn from a uniform distribution within the aforementioned ranges of values.

Figure \ref{fig:decorrelation} (left) shows the results of these calculations (Case A). We find angle differences between $-3.6^\circ$ and $1.95^\circ$, with the 5 and 95 percentiles of the distribution of the difference $\chi_{353} - \chi_{70}$ being -2.3$^\circ$ and 1.5$^\circ$, respectively. The angle difference depends slightly on the temperature difference, $\rm T^{IVC}_d - T^{LVC}_d$: for a given difference in $\beta$, the angle difference can vary up to $\sim$2$^\circ$. This is to be expected, since intensity is linearly proportional with temperature in the Rayleigh-Jeans limit. There is a much stronger dependence on the difference in spectral index, $\beta^{IVC}-\beta^{LVC}$: for a given temperature difference, the angle difference between frequencies can be as high as $\sim$5$^\circ$.

We investigate how the situation changes when we loosen our assumption that both clouds share a common ratio $P_{353}/p_R$ in Figure \ref{fig:decorrelation} (right), corresponding to Case B. We show the 5 and 95 percentiles of the distribution of $\chi_{353} - \chi_{70}$, found by evaluating equation \ref{eqn:deltachi} $10^3$ times for each of the 5 values of $r_{353}$ from Table \ref{tab:r353}. When the $P_{353}/p_R$ is such that it reduces the contribution of the IVC to the total emission ($r_{353} < 0.1$), we find negligible values for the angle difference. At higher $r_{353}$, the relative contribution of the IVC is increased and this is reflected in the distribution of angle differences. 

Our results show that for the observed region, the difference between the polarization angle measured at 353 GHz and that at 70 GHz will be at most $\sim 8^\circ$, if the IVC and LVC have properties within the assumed parameter ranges. 

\subsection{Other estimates for the distance to the LVC and IVC}

In section \ref{sec:dist} we found the distance to the LVC to be $\sim 360$ pc. We wish to compare this estimate with independent existing data. The 3D dust extinction map produced by \citet{green2015,green2018} using a Bayesian method on Pan-STARRS1 \citep{kaiser} data allows us to do this. 

\begin{figure}
\centering
\includegraphics[scale=1]{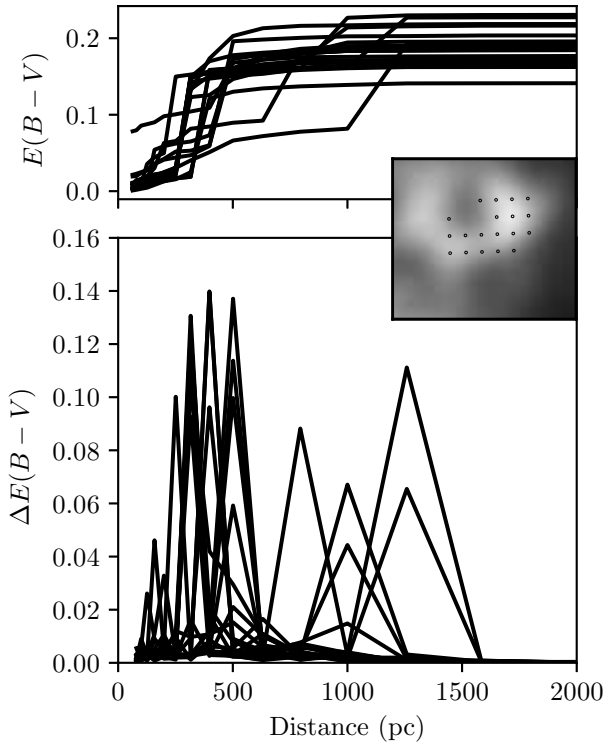}
\caption{Indications for two dust components in the three-dimensional stellar reddening map of \cite{green2018}. The inset marks the selected sightlines on the integrated emission of the IVC HI component.  Top panel: `Best-fit' E(B-V) as a function of distance for a number of sightlines towards the IVC cloud. Bottom panel: Differential reddening ($\rm \Delta E(B-V)$) for the same sightlines. }
\label{fig:dust3d}
\end{figure}

We selected 19 sightlines, spaced regularly every 10$\arcmin$ (comparable to the typical resolution of dust map at these latitudes) in Galactic $\ell$ and b, to cover the area of the IVC. As the method used by \citet{green2018} is probabilistic, we must choose an estimator to probe the  E(B-V) as a function of distance. We select to show the 'best-fit' estimate for each pixel. The uncertainty is captured by sampling different sightlines. The results are shown in Figure \ref{fig:dust3d}. The method provides a minimum reliable distance, after which there are enough stars to make a statistically significant inference. This varies within the sightlines selected. The maximum value found is 400 pc. On the other end of the distance scale, the maximum reliable distance is farther than 5000 pc. 

The presence of a cloud appears as a step in the E(B-V) versus distance curve (Fig. \ref{fig:dust3d}, top panel). As the reddening of this sightline is quite low, there is significant variation between the selected sightlines. However, the majority of sightlines agree on the position of the first such step. In the bottom panel, we show the numerical derivative of the E(B-V) curve, $\rm \Delta E(B-V)$. A step will appear as a peak in this plot. There is a clear over-density of high-$\rm \Delta$E(B-V) peaks at $\sim300$ pc. This is most likely the signature of the LVC and agrees well with our estimate of the distance to the cloud. The peak E(B-V) is found to be 0.12-0.14 mag, consistent with our upper limit of 0.16 in the 2-Cloud region (Table \ref{tab:NH}).

In 9 out of 19 sightlines, secondary peaks are evident. These, however, do not agree on the magnitude or distance of the dust component they are probing. Since the IVC has a very low reddening of 0.02-0.03 mag, this is comparable with the 25 mmag uncertainty on the optical reddening values \citet{schlafly2014}. The existence of a secondary peak in many of the sightlines supports the existence of the IVC, even if the exact properties of the cloud cannot be pinpointed. From these sightlines, it appears that the IVC most likely does not lie farther than $\sim1500$ pc (where there are no peaks observed). This is consistent with our estimation of the IVC distance based on the distance where the maximum $SNR\langle p \rangle_{IVC}$ is found ([1250 - 2140] pc, Section \ref{ssec:distance}).

\section{Summary} \label{sec:conclusions}

In this work we have demonstrated the technique of tomographic decomposition of the plane-of-the-sky magnetic field using precise starlight polarization measurements in combination with stellar distances inferred from \textit{Gaia}. For this demonstration, we selected a region towards the diffuse ISM which contains two distinct clouds along the line of sight (as evidenced by HI emission). We have tailored our experiment so that our starlight polarization traces not only the region with two clouds, but also a control region in which only one cloud is expected to  produce a signal. 

With a combination of diverse datasets, we are able to constrain a number of properties of the clouds. The local cloud lies at a distance of 346-393 pc, has a mean fractional linear polarization of 1.65$\pm$ 0.04 \% and a mean polarization angle of -27$^\circ \pm$ 1$^\circ$ and causes a mean reddening $E(B-V) \leq 0.16-0.18$ mag. The far cloud is located at a distance 1250$-$2140 pc, has a mean fractional linear polarization of 0.28$\pm$0.08\%, a mean polarization angle of 36$^\circ\pm$ 8$^\circ$, and E(B-V)$\geq 0.02$ mag. 

We have presented a new method of estimating the distance to the far cloud in this region, based on the dependence of the SNR of the mean fractional linear polarization on distance. We have evaluated the accuracy of the method in recovering the true distance and polarization properties of the far cloud.

Finally, we note that the stark differences between the properties of the two clouds pose a challenge to the task of decomposing the magnetic field along the line of sight. The local cloud dominates the signal (in both extinction and polarization) making it impossible to distinguish the effect of the farther cloud by simple inspection of the measurements as a function of distance. 
By providing a significant detection of the polarization of the farther cloud, we demonstrate that our method performs well even in this particularly difficult situation.






\acknowledgments

We thank A. Brimis, I. Komis, I. Leonidaki and N. Mandarakas for their help during the observations and T. Ghosh for helping with the analysis of Planck data. We thank N. Kylafis for useful suggestions on the paper. We thank the anonymous reviewer for their helpful comments. ANR, GVP and ACSR acknowledge support from the National Science Foundation, under grant number AST-1611547. This project has received funding from the European Research  Council  (ERC)  under  the  European  Union’s  Horizon  2020  research and innovation programme under grant agreement No. 771282. This work has made use of data from the European Space Agency (ESA) mission {\it Gaia} (\url{https://www.cosmos.esa.int/gaia}), processed by the {\it Gaia} Data Processing and Analysis Consortium (DPAC,  \url{https://www.cosmos.esa.int/web/gaia/dpac/consortium}). 
This research made use of the following python packages: APLpy \citep{aplpy}, Astropy \citep{astropy} and healpy \citep{healpix}.



\vspace{5mm}
\facilities{Skinakas:1.3m, Gaia, Effelsberg, CDS, IRSA, Planck}
\software{Astropy, aplpy, healpy}



\appendix

\section{SNR$_{\langle p \rangle}$ as a probe of the distance to the far cloud}
\label{sec:appendix}

In Section \ref{sec:tomography} we inferred the polarization properties of the near (LVC) and far (IVC) clouds averaged over stellar ensembles. We found the ensemble average Stokes parameters $\langle q \rangle$, $\langle u \rangle$ resulting from the effect of each cloud (separately) on starlight and calculated the mean fractional linear polarization $\langle p \rangle$ and polarization angle $\langle \theta \rangle$ with their associated uncertainties. Since we had no knowledge of the distance to the far cloud, this was performed for different assumed distances, $d^{IVC}$. As discussed in Section \ref{ssec:distance}, one expects that the maximal confidence in the measurement of the polarization properties of the far cloud should be obtained when the assumed $d^{IVC}$ coincides with the true distance to the cloud. 

\begin{figure*}
\centering
\includegraphics[scale=0.5]{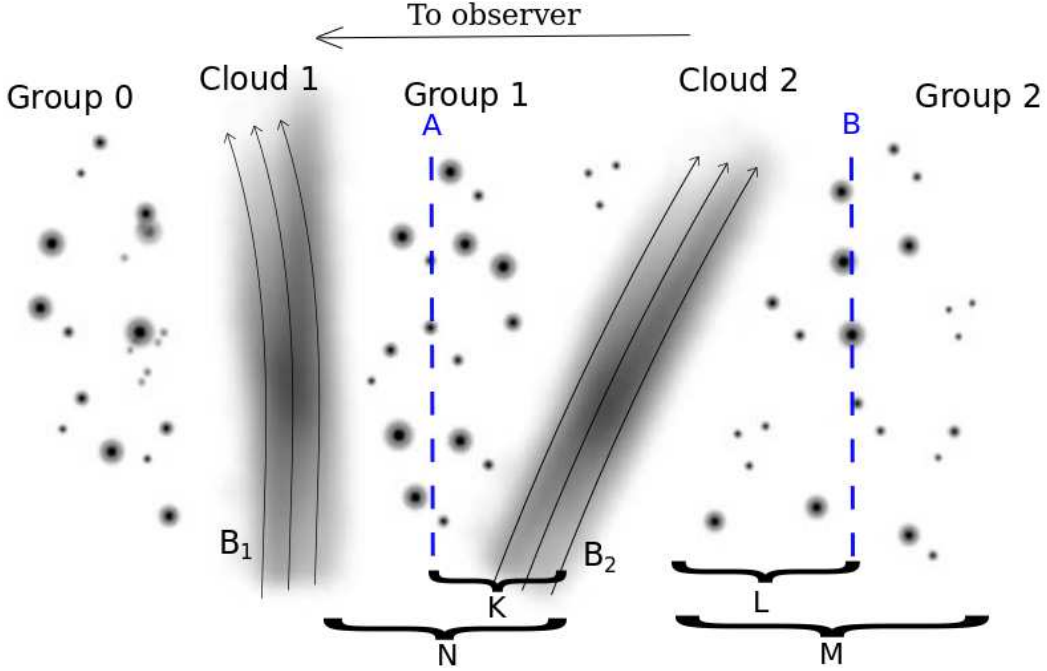}
\caption{Schematic of the distribution of stars towards a line of sight with 2 clouds (Cloud 1, Cloud 2). The observer lies towards the left edge of the figure. Each star belongs to one of three groups according to its position relative to the clouds. Vertical dashed lines mark the two cases discussed in the appendix: A, the assumed distance to Cloud 2 is less than the true distance and B, the assumed distance is larger than the true distance. Horizontal brackets in the bottom are labelled by the number of stars in the corresponding distance range. The magnetic fields of the two clouds are shown with smooth black lines, marked B1 and B2.}
\label{fig:cartoon}
\end{figure*}

In this appendix we support this intuitive picture with a simplistic mathematical proof. We consider the case of two clouds lying along the line of sight, as illustrated by the cartoon in Figure \ref{fig:cartoon}. The cloud that is nearest to the observer, Cloud 1 in Figure \ref{fig:cartoon}, is denoted as `$C1$' and that which is farther away, Cloud 2 in Figure \ref{fig:cartoon}, is denoted as `$C2$'. Stars belong to three groups: Group 0 (foregrounds), Group 1 (between $C1$ and $C2$) and Group 2 (backgrounds). 

We begin by making a few simplifying assumptions to facilitate the calculations. First, we assume that the ensemble averages ($\langle q \rangle^{C1}$, $\langle u \rangle^{C1}$) found using stars in Group 1 are a good descriptor of the distribution of q, u generated by Cloud 1, so that $\langle q \rangle^{C1} = q^{C1}$ (and similarly for $u^{C1}$). Similarly, for stars in Group 2, we assume that $\langle q \rangle^{C1} + \langle q \rangle^{C2} = q^{C1} + q^{C2}$ (equivalently for $u^{C1}+u^{C2}$).
Second, we make the following assumptions for the measurement uncertainties. Due to the way that stellar polarizations are measured, a usually valid approximation is that the uncertainty of each stellar measurement in $q$ is equal to that in $u$ so that: $\sigma_{q,i} = \sigma_{u,i} = \sigma_i$. This is the case for our data as well, with 90\% of measurements having $|\sigma_{q,i} - \sigma_{u,i}| < 0.1\%$. 
We will be using the common average instead of the weighted average of stars in Groups 1 and 2 for the following calculations, as this facilitates the interpretation of the final result. The implicit assumption here is that all stellar measurements are equal, i.e. $\sigma_i = \sigma$. This seems as a rather crude approximation: our measurement uncertainties in $q$ and $u$ are distributed with a mean of $0.46\%$ and a standard deviation of $0.2\%$. However, the error that we make with this assumption is insignificant, as the mean for both groups is at the level of 1\%. As a result of the aforementioned assumptions we obtain:
\begin{equation}
\left\langle q\right\rangle^{Group} = \sum_{i=1}^{N} \frac{q_i}{N}, \left\langle u\right\rangle^{Group} = \sum_{i=1}^{N} \frac{u_i}{N}, 
\label{eq:means}
\end{equation}
where N is the number of stars in the group under consideration.

We now wish to investigate how the assumed distance to Cloud 2 affects the polarizing properties we infer for this cloud if we follow the process of decomposition outlined in Section \ref{sec:tomography}. We consider the following two cases: (A) The assumed distance is less than the true distance (e.g. left vertical dashed line in Figure \ref{fig:cartoon}) and (B) the assumed distance is larger than the true distance (e.g. right vertical dashed line in Figure \ref{fig:cartoon}). 

\subsection{Case A: Assumed distance to Cloud 2 is less than true distance}
In this case, a number K of stars which in reality lie in Group 1 will be erroneously assigned to Group 2. The mean properties we find for Cloud 1 will be: 
\begin{equation}
\left\langle q \right\rangle^{C1,A} = \sum_{i=1}^{N-K} \frac{q_i}{N-K}
\label{eq:C1A}
\end{equation}
(and similarly for $u$). For Cloud 2 we will find:
\begin{equation}
\left\langle q \right\rangle^{C1+2,A} = \sum_{i=1}^{M+K} \frac{q_i}{M+K} \,\,\,\, \Rightarrow \,\,\,\, \left\langle q \right\rangle^{C1+2,A} = \sum_{i=1}^{K} \frac{q_i}{M+K} + \sum_{i=1}^M \frac{q_i}{M+K}
\label{eq:C2A}
\end{equation}
From equation \ref{eq:means} we find $\sum_{i=1}^{K} {q_i} = K \left\langle q\right\rangle^{C1}$. Note that if K is small, the ensemble average will not necessarily equal the true $\left\langle q\right\rangle^{C1}$. However, in this case the effect of the first term in equation \ref{eq:C2A} will not be significant compared to the second term which will arise from a much larger number of stars. For Group 2 stars we will have $\sum_{i=1}^{M} q_i = M(\left\langle q\right\rangle^{C1} + \left\langle q\right\rangle^{C2})$. Substituting these two expressions into equation \ref{eq:C2A}, we find:
\begin{equation}
\left\langle q \right\rangle^{C1+2,A} = \frac{K\left\langle q\right\rangle^{C1}}{M+K} + \frac{M(\left\langle q\right\rangle^{C1}+\left\langle q\right\rangle^{C2})}{M+K}
\label{eq:C2A_2}
\end{equation}
Next, we calculate the mean Stokes parameters of Cloud 2 only, as in Section \ref{sec:tomography}:
\begin{equation}
\left\langle q\right\rangle^{C2,A} = \left\langle q\right\rangle^{C1+2,A} - \left\langle q\right\rangle^{C1,A} \Rightarrow 
\left\langle q\right\rangle^{C2,A} = \frac{M}{M+K} \left\langle q\right\rangle^{C2},
\end{equation}
where we have used equations \ref{eq:C1A} and \ref{eq:C2A_2}.
It is easy to see that when the distance to Cloud 2 is chosen correctly (K=0), we recover the correct value of $\left\langle q\right\rangle^{C2}$. In fact, when this is the case the final expression obtains its maximum (absolute) value.

\subsection{Case B: Assumed distance to Cloud 2 is larger than true distance}
Next, we repeat the analysis for Case B, when the assumed distance to Cloud 2 is larger than the true distance. In this case there are L stars from Group 2 mis-attributed to Group 1. Following the same reasoning as in case A, we obtain for the mean of Cloud 1: 
\begin{equation}
\left\langle q\right\rangle^{C1,B} = \sum_{i=1}^{N+L} \frac{q_i}{N_L} \Rightarrow \left\langle q\right\rangle^{C1,B} \sum_{i=1}^{N+L} \frac{q_i}{N+L} + \sum_{i=1}^{L} \frac{q_i}{N+L} \Rightarrow \left\langle q\right\rangle^{C1,B} = \frac{N}{N+L}\left\langle q\right\rangle^{C1} + \frac{L}{N+L} (\left\langle q\right\rangle^{C1}+\left\langle q\right\rangle^{C2})
\end{equation}
The mean of Group 2 stars will simply be: 
\begin{equation}
\left\langle q\right\rangle^{C1+2,B} = \sum_{i=1}^{M-L} \frac{q_i}{M-L}
\end{equation}
We shall assume here that the average of a random subsample of Group 2 is equal to the average of the whole sample. This assumption of course will break down in the limit of small numbers. But in this case, M-L=0 and there are very few stars left to evaluate the properties of Cloud 2. With this assumption, we can rewrite the previous equation as:
\begin{equation}
\left\langle q\right\rangle^{C1+2,B} = \left\langle q\right\rangle^{C1} + \left\langle q\right\rangle^{C2}
\end{equation}
Finally, we subtract the effect of Cloud 1 to obtain the mean properties of Cloud 2:
\begin{equation}
\left\langle q\right\rangle^{C2,B} = \left\langle q\right\rangle^{C1+2,B} - \left\langle q\right\rangle^{C1,B}
\Rightarrow \left\langle q\right\rangle^{C1+2,B} = \left( 1-\frac{L}{N+L}\right) \left\langle q\right\rangle^{C2}
\end{equation}
Once again, the correct value is of course recovered when the assumed distance is equal to the true distance, hence L=0. But also, it is plain to see that the expression reaches a maximum when this happens. In summary, we have found that the mean Stokes parameters of Cloud 2 achieve their maximum (absolute) values when the assumed distance to the cloud is the correct one.

\subsection{SNR$\left\langle p\right\rangle^{IVC}$ as a function of assumed cloud distance}
The average Stokes parameters are expected to reach their maximum (in absolute value) at the true distance to the IVC. It follows from equation \ref{eq:polarization_stokes} that the same will hold for the fractional linear polarization. The associated uncertainties on these values vary by less than 0.06\% throughout the range of assumed distances. As a result, the maximum of the SNR in $p$ is expected to lie at the true distance of the cloud (within our sampling error of 200-300 pc). 

In practice, however, we must (and do) include the uncertainties of the measurements in the calculation of the Stokes parameters for the IVC. The weighted average is not necessarily maximal at the same distance as the unweighted one used to derive the previous expressions. It is the weighted averages ($\left\langle q\right\rangle^{IVC}$, $\left\langle u\right\rangle^{IVC}$) that go into the calculation of $\left\langle p\right\rangle^{IVC}$. In addition to this, we have ignored the effect of bias on the $\left\langle p\right\rangle^{IVC}$, which could have an effect on the location of the maximum $SNR$ if $\left\langle p\right\rangle^{IVC}$ were not significantly detected. Section \ref{ssec:distance} evaluates the effectiveness of the maximum SNR$\left\langle p\right\rangle_d^{IVC}$ in detecting the true distance to the cloud without the simplifying assumptions made in this section.

\section{Estimation of reddening for the IVC and LVC}
\label{sec:appendixB}

In this appendix we derive estimates of the mean reddening caused by the IVC and LVC in both observed regions, used in section \ref{ssec:pebv}.

In the 2-Cloud region, the IVC has $\rm N_{HI}$ of $\sim 2\times$10$^{20}$cm$^{-2}$, which corresponds to the transition from atomic to molecular hydrogen \citep[e.g.][]{gillmon}. Consequently, the IVC in this region may contain small amounts of $\rm H_2$. Evidence for this comes from the fact that the western part of the IVC partially overlaps with cloud number 141 in the catalogue of candidate molecular-IVCs from \citet{roehser} (l= 103.6$^\circ$, b=22.5$^\circ$). With a molecular fraction $f_{mol} = \rm 2N^{IVC}_{H_2}/N^{IVC}_H = E(B-V)^{IVC}_{H2}/E(B-V)^{IVC}$ (the ratio of the reddening due to the molecular component over the total reddening of the IVC), we can write the reddening of the IVC as:
\begin{equation}
{\rm E(B-V)^{IVC}  = E(B-V)^{IVC}_{HI}} \frac{1}{1 - f_{mol}}, 
\end{equation}
where $\rm E(B-V)^{IVC}_{HI}$ is the E(B-V) derived from converting the $\rm N^{IVC}_{HI}$ to reddening. To this end, we use the relation from \cite{Lenz}, which holds for $\rm N_{HI} < 4 \times 10^{20}  cm^{-2}$: $E(B-V) = N_{HI}/(8.8\times10^{21})\rm mag/cm^{-2}$. The molecular IVCs in the Northern hemisphere sample of \citet{roehser}, have low molecular fractions (median $f_{mol}\sim$0.5). For lack of additional information on the specific IVC, we choose to place only a lower limit on the reddening of this cloud, given by $f_{mol}=0$. Thus, we obtain for the IVC in the 2-Cloud region: $\rm E(B-V)^{IVC} > 0.02 \rm mag$. The typical scatter in the conversion from $\rm N_{HI}$ to $\rm E(B-V)$ is 5 mmag and is therefore negligible compared to the uncertainty introduced by $f_{mol}$.

With $\rm N^{LVC}_{HI} =  3.5 \times 10^{20}$cm$^{-2}$ (3.6 $\times 10^{20}$cm$^{-2}$ in the 1-Cloud region), the LVC most likely contains a significant amount of molecular material. The reddening caused by the LVC alone will therefore be:
\begin{equation}
\rm E(B-V)^{LVC} = E(B-V)^{LVC}_{H_2} + E(B-V)^{LVC}_{HI},
\label{eqn:ebvlvc}
\end{equation}
where $\rm E(B-V)^{LVC}_{H_2}$ is the reddening that arises from the molecular component of the LVC. We can use the total reddening, inferred for example from thermal dust emission ($\rm E(B-V)_d$), to estimate the reddening that arises from molecular gas throughout the sightline, $\rm E(B-V)_{H_2}$:
\begin{equation}
\rm E(B-V)_{H_2} = E(B-V)_d - E(B-V)_{HI}
\label{eqn:ebvh2}
\end{equation}
where all the values refer to material integrated over the entire line of sight. We note here that in equation \ref{eqn:ebvh2}, we assume that all the material that is not traced by HI is molecular. Hence we ignore the effect of optically thick HI emission \citep[as shown by][this is a valid assumption for the local ISM]{murray2018}.

We obtain the total extinction $A_V$ from the map presented in \cite{planckxxix} and convert to reddening ($\rm E(B-V)_d$) assuming a ratio of total to selective extinction of $R_V = 3.1$. The map has a pixel size of 1.7\arcmin (sampled on a HEALPix grid of NSIDE 2048) and we use the average extinction within a circular disk centred on both regions with a radius of 0.16$^\circ$. We find $\rm E(B-V)_d = 0.21 \,\rm mag$ in the 2-Cloud region (0.23 mag in the 1-Cloud region). Therefore, the $\rm E(B-V)_{HI}$ of 0.09 mag (integrated over all velocities) accounts for less than half the total reddening of the sightline. 

The remaining reddening must arise from the HI-dark (molecular) material. This material is certainly not associated with the diffuse HI emission that is not part of the IVC or LVC components. While the column density associated with this emission is comparable to that of the LVC, its source is highly spread out in (velocity) space so that significant shielding from the radiation field (necessary for the creation of molecular hydrogen) cannot be attained. 
If the IVC has zero $f_{mol}$, then $\rm E(B-V)^{LVC}_{H_2} = E(B-V)_{H_2}$ and we obtain an upper limit on the reddening of the LVC by substituting from equation \ref{eqn:ebvh2} into equation \ref{eqn:ebvlvc}: $\rm E(B-V)^{LVC} \leq E(B-V)_{H_2} + E(B-V)^{LVC}_{HI}$. The resulting values for both regions are shown in Table \ref{tab:NH}. 

In the 1-Cloud region, we can better constrain the reddening of the two components, as the IVC exhibits too low a column density ($\rm N^{IVC}_{HI} =  0.9 \times 10^{20}$cm$^{-2}$) to harbor a significant amount of $\rm H_2$. We can attribute the entirety of the molecular material in this sightline to the LVC. From the results, shown in Table \ref{tab:NH}, we deduce that the LVC has a molecular fraction of $f^{LVC}_{mol} = 0.75$. This is in agreement with other LVCs at similar total column densities (1.6$\times 10^{21}$cm$^{-2}$) found in the study of \citet{planckdusthalo} (their Fig. 20). 
The molecular fraction found in the 2-Cloud region (using the upper limit on E(B-V)$^{LVC}$) is only 1\% lower than that found in the 1-Cloud region.

\section{Dust emission polarization angle in the case of two clouds}
\label{sec:appendixC}

In this Appendix we derive the expression used in Section \ref{ssec:decorrelation} for the polarization angle of thermal dust emission in the case of two components (clouds) lying along the line of sight\footnote{Our derivation differs from that of \citet{tassis} in that we do not assume the same spectral index for both clouds and we use the ratio of polarized intensities of the two clouds instead of the ratio of total intensities.}. The total intensity of cloud $C_i$ at frequency $\nu$ ($I^{C_i}_\nu$, where $i$ = 1,2) is modelled as a modified gray-body, following e.g. \citet{planckxi_thermal}:

\begin{equation}
I^{C_i}_\nu \propto c^{C_i} (\frac{\nu}{\nu_0})^{\beta^{C_i}} N^{C_i}_H B(\nu,T^{C_i}),
\label{eqn:Iv}
\end{equation}
where $c^{C_i}$ is the dust-to-gas mass ratio in the cloud, $\nu_0$ is a reference frequency, $\beta^{C_i}$ is the spectral index of the power-law dust emissivity, ${\rm N}^{C_i}_{\rm H}$ the cloud hydrogen column density, and $B(\nu,T^{C_i})$ the Planck function for dust at temperature $T^{C_i}$.

The Stokes parameters $Q^{C_i}_\nu$ and $U^{C_i}_\nu$ for cloud ${C_i}$ at frequency $\nu$ are given by:

\begin{equation}
Q^{C_i}_\nu = p^{C_i}_\nu I^{C_i}_\nu \cos{2\chi^{C_i}}, \,\,\, U^{C_i}_\nu = p^{C_i}_\nu I^{C_i}_\nu \sin{2\chi^{C_i}}\footnote{The angle $\chi$ is measured according to the IAU convention.},
\label{eqn:QUemission}
\end{equation}
where $p^{C_i}_\nu$ is the fractional linear polarization of cloud ${C_i}$ at frequency $\nu$ ($p^{C_i}_\nu = \sqrt{(Q^{C_i}_\nu)^2+(U^{C_i}_\nu)^2}/I^{C_i}_\nu$), and $\chi^{C_i}$ is the polarization angle of the emission (which depends only on the plane-of-sky orientation of the magnetic field threading the cloud and therefore does not have a frequency dependence). 

The emission reaching the observer will have Stokes parameters given by the sum of the signals coming from both clouds:
\begin{equation}
Q_\nu = Q^{C_1}_\nu + Q^{C_2}_\nu, \,\,\, U_\nu = U^{C_1}_\nu + U^{C_2}_\nu , 
\label{eqn:QUsum}
\end{equation}

and so the polarization angle observed will be:
\begin{equation}
\chi_\nu = \frac{1}{2} \arctan{\frac{U_\nu}{Q_\nu}}  \Rightarrow 
\chi_\nu  =  \frac{1}{2} \arctan{\frac{p^{C_1}_\nu I^{C_1}_\nu \sin{2\chi^{C_1}} + p^{C_2}_\nu I^{C_2}_\nu \sin{2\chi^{C_2}}}{p^{C_1}_\nu I^{C_1}_\nu \cos{2\chi^{C_1}} + p^{C_2}_\nu I^{C_2}_\nu \cos{2\chi^{C_2}}}}
\label{eqn:chinu}
\end{equation}

We define the ratio of the polarized intensities ($P = \sqrt{Q^2 + U^2}$) of the two clouds as:
\begin{equation}
r_{\nu} = \frac{P^{C_1}_{\nu}}{P^{C_2}_{\nu}} = \frac{p^{C_1}_{\nu} I^{C_1}_{\nu}}{p^{C_2}_{\nu} I^{C_2}_{\nu}},
\label{eqn:ratioP}
\end{equation}
and use this to re-write equation \ref{eqn:chinu} as:
\begin{equation}
\chi_\nu = \frac{1}{2} \arctan{ \frac{r_{\nu} \sin{2\chi^{C_1}} + \sin{2\chi^{C_2}} }{ r_{\nu} \cos{2\chi^{C_1}} + \cos{2\chi^{C_2}} }}.
\label{eqn:chinur}
\end{equation}

It is now plain to see from equation \ref{eqn:chinur}, that the difference between the polarization angle at two frequencies $\nu_1$, $\nu_2$ depends on the parameters: $\chi^{C_1}, \chi^{C_2}, r_{\nu_1}, r_{\nu_2}$. Since we have measured the plane-of-sky magnetic field orientation in each cloud (Section \ref{sec:tomography}), the first two parameters are known: 
\begin{equation}
\chi^{C_1} = \theta_{LVC} + 90^\circ, \,\,\, \chi^{C_2} = \theta_{IVC} + 90^\circ .
\end{equation}
The parameters $r_{\nu_1}$ and $r_{\nu_2}$ depend on $p^{C_1}_{\nu}, p^{C_2}_{\nu}, T^{C_1}, T^{C_2}, \beta^{C_1}, \beta^{C_2}$. We can use supplementary information to reduce the number of free parameters. Measurements from \textit{Planck} \citep{planckpconst} and BLASTPol \citep{ashton} show that $p_\nu$ is constant for a wide range of frequencies. Therefore we will take $p^{C_i}_{\nu_1} = p^{C_i}_{\nu_2}$. Since $p_\nu = P_\nu/I_\nu$, equation \ref{eqn:ratioP} becomes:
\begin{eqnarray}
r_{\nu_2} &=& \frac{P^{C_1}_{\nu_2}}{P^{C_2}_{\nu_2}} \Rightarrow r_{\nu_2}=  \frac{P^{C_1}_{\nu_1} \frac{I^{C_1}_{\nu_2}}{I^{C_1}_{\nu_1}}}{P^{C_2}_{\nu_1} \frac{I^{C_2}_{\nu_2}}{I^{C_2}_{\nu_1}}} \Rightarrow 
r_{\nu_2}  = r_{\nu_1} \frac{I^{C_1}_{\nu_2} I^{C_2}_{\nu_1}}{ I^{C_2}_{\nu_2} I^{C_1}_{\nu_1}} \Rightarrow \nonumber \\
r_{\nu_2}  &=& r_{\nu_1} \frac{B(\nu_2,T^{C1})}{B(\nu_1,T^{C_2})} (\frac{\nu_2}{\nu_1})^{\beta^{C_1}} \frac{B(\nu_2,T^{C_2})}{B(\nu_1,T^{C_2})} (\frac{\nu_1}{\nu_2})^{\beta^{C_2}},
\end{eqnarray}
where we have made use of equation \ref{eqn:Iv}, under the assumption that the gas-to-mas ratio between the two clouds is the same ($c^{C_1} = c^{C_2}$).

\listofchanges


\begin{thebibliography}{}

\bibitem[Andersson \& Potter (2005)]{anderssonpotter} Andersson B. G.; Potter S. B.; 2005; MNRAS; 356; 1088
\bibitem[Appenzeller (1968)]{appenzeller} Appenzeller I., 1968, ApJ, 151, 907
\bibitem[Astropy Colaboration (2013)]{astropy} Robitaille T. P. et al.; 2013; A\&A; 558A; 33A
\bibitem[Ashton et al. (2018)]{ashton} Ashton P. C.; et al.; 2018; ApJ; 857; 10
\bibitem[Bailer-Jones et al. (2018)]{bailer-jones} Bailer-Jones, C. A. L.; Rybizki, J.; Fouesneau, M.; Mantelet, G.; Andrae, R.; 2018; AJ 156; 58
\bibitem[Berdyugin \& Teerikorpi (2002)]{Berdyugin2002} Berdyugin A.; Teerikorpi P.; 2002; A\&A; 384; 1050
\bibitem[Berdyugin et al.(2014)]{berdyugin} Berdyugin, A., Piirola, V., Teerikorpi, P., 2014, A\&A, 561, 24
\bibitem[Bohlin et al. (1978)]{Bohlin} Bohlin, R. C., Savage, B. D., \& Drake, J. F. 1978, ApJ, 224, 132
\bibitem[Breger (1986)]{breger1986} Breger, M.; 1986; ApJ; 309; 311B
\bibitem[Breger (1987)]{breger1987} Breger, M.; 1987; ApJ; 319; 754B
\bibitem[Boulanger et al. (2018)]{imagine} Boulanger F. et al., 2018, JCAP, 08, 049B
\bibitem[Chen et al. (2017)]{chen} Chen Z.; Jiang, Z.; Tamura, M.; Kwon, J.; Roman-Lopes, A.; 2017; ApJ; 838; 80
\bibitem[Clark et al. (2014)]{clark} Clark, S. E.; Peek, J. E. G.; Putman, M. E.; 2014; ApJ; 789; 82C
\bibitem[Clemens et al. (2012)]{clemens} Clemens, Dan P.; Pinnick, A. F.; Pavel, M. D.; Taylor, B. W.; 2012; ApJS; 200; 19C
\bibitem[Cotton et al. (2016)]{cotton} Cotton D. V.; et al.; 2016; MNRAS; 455; 1607
\bibitem[Coyne (1974)]{coyne} Coyne G. V.; 1974; AJ; 79; 565
\bibitem[Ellis \& Axon (1978)]{ellisaxon1978} Ellis, R. S.; Axon, D. J.; 1978; Ap\&SS; 54; 425E
\bibitem[Eswaraiah et al. (2012)]{eswaraiah2012}	Eswaraiah, C.; Pandey, A. K.; Maheswar, G.; Chen, W. P.; Ojha, D. K.; Chandola, H. C.; 2012; MNRAS; 419; 2587E
\bibitem[Ferri\`ere (2016)]{ferriere} Ferri\`ere, K., 2016, JPhCS, 767, 2006F
\bibitem[Fowler \& Harwit (1974)]{fowler1974} Fowler, L. A.; Harwit, M.; 1974; MNRAS; 167; 227F
\bibitem[Franco \& Alves (2015)]{franco} Franco G. A. P.; Alves F. O.; 2015; ApJ; 807; 5F
\bibitem[Gaia Collaboration (2016)]{gaiamission} Prusti, T. et al., Gaia Collaboration; 2016; A\&A; 595A; 1G
\bibitem[Gaia Collaboration (2018)]{gaiadr2} Brown, A. G. A.; Vallenari, A.; Prusti, T.; de Bruijne, J. H. J.; Babusiaux, C.; Bailer-Jones, C. A. L.; Gaia Collaboration; 2018; A\&A; 616A; 1G
\bibitem[Gillmon et al. (2006)]{gillmon} Gillmon, K.; Shull, J. M.; Tumlinson, J.; \& Danforth, C.; 2006; ApJ; 636; 891
\bibitem[Gorski et al. (2005)]{healpix} G\'{o}rski, K. M.; et al.; 2005; ApJ; 622; 759G
\bibitem[Green et al. (2015)]{green2015} Green G. M., et al.; 2015; ApJ; 810; 25 
\bibitem[Green et al. (2018)]{green2018} Green G. M. et al.; 2018; MNRAS; 478; 651G
\bibitem[Heald et al. (2015)]{heald} Heald G., Beck R., de Blok W.J.G. et al., 2015, AASKA14, 106
\bibitem[Heiles (2000)]{heiles} Heiles, C.; 2000; AJ; 119; 923
\bibitem[Hensley \& Bull (2018)]{hensley} Hensley, B.; Bull, P.; 2018; ApJ; 853; 127H
\bibitem[HI4PI Collaboration (2016)]{HI4PI} HI4PI Collaboration; 2016; A\&A; 594A; 116H
\bibitem[Hiltner (1956)]{hiltner} Hiltner, W. A.; 1956; ApJS; 2; 389H
\bibitem[Kaiser et al. (2010)]{kaiser} Kaiser N. et al; SPIE; 7733; 77330E
\bibitem[King et al. (2014)]{King2014} King, O.G.; et al.; 2014; MNRAS; 442; 1706
\bibitem[Kulkarni \& Heiles (1988)]{kulkarni} Kulkarni, S. R., and Heiles, C. 1988, in Galactic and Extragalactic Radio Astronomy, 2nd edition, eds. G. L. Verschuur and K. I. Kellermann, (New York: Springer-Verlag), p. 95.
\bibitem[Lee \& Draine (1985)]{LeeDraine} Lee H. M.; Draine B. T.; 1985, ApJ, 290, 211
\bibitem[Lenz, Hensley \& Dor\'e (2017)]{Lenz} Lenz, D., Hensley, B. S. \& Dor\'{e}, O., 2017, ApJ, 846, 38
\bibitem[Li et al. (2006)]{Li2006} Li H. et al.; 2006; ApJ; 648; 340
\bibitem[Lloyd \& Harwit (1973)]{lloyd1973} Lloyd, S.; Harwit, M. O.; 1973; IAUS; 52; 203L
\bibitem[Magalh\~{a}es et al. (2012)]{southpol} Magalh\~aes A. M., et al., 2012, AIP Conf. Ser. 1429, ed. J. L. Hoffman,
J. Bjorkman, \& B. Whitney (New York: AIP), 244
\bibitem[Marchwinski, Pavel \& Clemens (2012)]{marchwinski} Marchwinski, R. C.; Pavel, M. D.; Clemens, D. P.; 2012; ApJ; 755; 130
\bibitem[Martin (1974)]{martin} Martin P. G.; 1974; ApJ; 187; 461
\bibitem[Monet et al. (2003)]{usnob} Monet, D. G.; 2003; AJ; 125; 984M
\bibitem[Montier et al. (2015a)]{montiera} Montier, L.; Plaszczynski, S.; Levrier, F.; Tristram, M.; Alina, D.; Ristorcelli, I.; Bernard, J.-P.; 2015a; A\&A; 574; A135
\bibitem[Montier et al. (2015b)]{montier} Montier, L.; Plaszczynski, S.; Levrier, F.; Tristram, M.; Alina, D.; Ristorcelli, I.; Bernard, J.-P.; Guillet, V.; 2015b; A\&A; 574; A136
\bibitem[Murray et al. (2018)]{murray2018} Murray C. E.; Peek, J. E. G.; Lee, M.-Y.; Stanimirovic, S.; 2018; ApJ; 862; 131M
\bibitem[Naghizadeh-Khouei \& Clarke (1993)]{Naghizadeh1993} Naghizadeh-Khouei, J.; Clarke, D.; 1993; A\&A; 274; 968N
\bibitem[Ostriker,  Stone,  \& Gammie (2001)]{ostriker} 	Ostriker E. C., Stone J. M., Gammie C. F., 2001, ApJ, 546, 980
\bibitem[Panopoulou et al. (2015)]{Panopoulou2015} Panopoulou G.; et al.; 2015; MNRAS; 452; 715
\bibitem[Panopoulou et al. (2016)]{erratum} Panopoulou G.; et al.; 2016; MNRAS; 462; 2011P
\bibitem[Patat et al.(2010)]{Patat2010} Patat, F.; Maund, J. R.; Benetti, S.; Botticella, M. T.; Cappellaro, E.; Harutyunyan, A.; Turatto, M., 	2010, A\&A, 510A; 108P
\bibitem[Pavel (2014)]{pavel} Pavel, M., D.; 2014; ApJ; 148; 49
\bibitem[Pereyra \& Magalh\~{a}es (2004)]{pereyra} Pereyra A., Magalh\~{a}es A. M.; 2004; ApJ; 603; 584
\bibitem[Planck Collaboration (2011)]{planckdusthalo} Planck Collaboration, 2011, A\&A, 536, A24
\bibitem[Planck Collaboration (2014a)]{planckxvii} Planck Collaboration, 2014a, A\&A, 566, A55
\bibitem[Planck Collaboration (2014b)]{planckxi_thermal} Planck Collaboration, 2014b, A\&A, 571, A11
\bibitem[Planck Collaboration (2015a)]{planckxix} Planck Collaboration, 2015a, A\&A, 576, A104
\bibitem[Planck Collaboration (2015b)]{planckpconst} Planck Collaboration, 2015b, A\&A, 576, A107
\bibitem[Planck Collaboration (2016a)]{planckmodel} Planck Collaboration, 2016a, A\&A, 596, A103
\bibitem[Planck Collaboration (2016b)]{planckalignment} Planck Collaboration, 2016b, A\&A, 586, A135
\bibitem[Planck Collaboration (2016c)]{planckxlviii} Planck Collaboration, 2016c, A\&A, 596, A109
\bibitem[Planck Collaboration (2016d)]{planckx} Planck Collaboration, 2016d, A\&A, 594, A10
\bibitem[Planck Collaboration (2016e)]{planckxxix} Planck Collaboration, 2016e, A\&A, 586, A132
\bibitem[Planck Collaboration (2018a)]{planckxii} Planck Collaboration, 2018a, A\&A, submitted, arXiv:1807.06212
\bibitem[Planck Collaboration (2018b)]{planckhfi} Planck Collaboration, 2018b, A\&A, accepted, AA/2018/32909, arXiv:1807.06207
\bibitem[Planck Collaboration (2018c)]{planckxi2018} Planck Collaboration, 2018c, A\&A, submitted, arXiv:1801.04945
\bibitem[Plaszczynski et al. (2014)]{plaszczynski} Plaszczynski, S.; Montier, L.; Levrier, F.; Tristram, M.; 2014; MNRAS; 439; 4048
\bibitem[Poh \& Dodelson (2017)]{poh} Poh, J.; Dodelson, S.; 2017; PhRvD; 95; 3511P
\bibitem[Ramaprakash et al. (2019, in prep)]{ramaprakash} Ramaprakash A. N.; et al.; 2019; in prep.
\bibitem[Robitaille \& Bressert (2012)]{aplpy} Robitaille, T.; Bressert, E.; 2012; APLpy: Astronomical Plotting Library in Python; Astrophysics Source Code Library, ascl:1208.017
\bibitem[Roehser et al. (2016)]{roehser} Roehser, T., Kerp, J., Lenz D., Winkel, B., 2016, A\&A, 596; A94
\bibitem[Santos, Corradi \& Reis (2011)]{santos2011} Santos, F. P.; Corradi, W.; Reis, W.; 2011; ApJ; 728; 104
\bibitem[Schlafly et al. (2014)]{schlafly2014} Schlafly E. F. et al.; 2014; ApJ; 789; 15S
\bibitem[Schmidt et al. (1992)]{schmidt} Schmidt G. D.; Elston R.; Lupie O. L.; 1992; AJ; 104; 1563
\bibitem[Serkowski (1962)]{serkowski1962} Serkowksi, K.; 1962; Adv. Astron. Astrophys.; 1; 289
\bibitem[Serkowski, Mathewson \& Ford (1975)]{serkowski} Serkowski, K.; Mathewson, D. S.; Ford, V. L., 1975, ApJ, 196, 261S
\bibitem[Skalidis et al. (2018)]{Skalidis2018} Skalidis R.; Panopoulou G. V.; Tassis K.; Pavlidou V.; Blinov D.; Komis I.; Liodakis I.; 2018; A\&A; 616A; 52S
\bibitem[Tassis \& Pavlidou (2015)]{tassis} Tassis K.; Pavlidou V.; 2015; MNRAS; 451, L90
\bibitem[Tassis et al. (2018)]{pasiphae} Tassis K.; et al.; 2018; arXiv:1810.05652
\bibitem[Treanor (1963)]{treanor1963} Treanor P. J.; 1963; AJ; 68; 185
\bibitem[Turnshek et al. (1990)]{Turnshek} {Turnshek}, D.~A. et al.; 1990; \aj ; 99; 1243
\bibitem[Vaillancourt (2006)]{vaillancourt} Vaillancourt, J. E.; 2006; PASP; 118; 1340V
\bibitem[Vergne et al. (2007)]{vergne} Vergne, M. M.; Feinstein, C.; Martínez, R.; 2007; A\&A; 462; 621V
\bibitem[Wakker (2001)]{wakker} Wakker, B. P.; 2001; ApJS; 136; 463

\end{thebibliography}
\end{document}